\newcommand{\CLASSINPUTbaselinestretch}{1}
\newcommand{\CLASSINPUTbottomtextmargin}{0.8in}
\documentclass[journal]{IEEEtran}
\usepackage{etoolbox}
\makeatletter
\patchcmd{\maketitle}{\@copyrightspace}{}{}{}
\makeatother

\usepackage[]{graphicx}
\usepackage{stfloats}  
\usepackage{cite}  
\usepackage{psfrag}  
\usepackage{subfigure}
\usepackage{wrapfig}
\usepackage{epsfig}
\usepackage{url}
\usepackage{amsmath, amssymb}
\usepackage{array}
\usepackage{multicol}
\usepackage{algorithm}
\usepackage{algorithmic}
\usepackage{mathrsfs}
\usepackage{multirow}
\usepackage[normalem]{ulem}

\usepackage[utf8x]{inputenc} 
\usepackage{ucs} 
\usepackage{amsfonts} 
\usepackage{makeidx} 
\usepackage[dvipsnames,usenames]{color} 
\usepackage{array} 
\usepackage{colortbl} 
\usepackage{booktabs}

\definecolor{tableheading}{rgb}{0.9,0.9,0.9}
\definecolor{softblue}{rgb}{0.8,0.8,1} 
\arrayrulecolor{ForestGreen} 
\usepackage{tabu}
\usepackage[justification=centering]{caption}

\begin{document}


\title{Computing in Memory with Spin-Transfer Torque Magnetic RAM\vspace{-0in}
\thanks{This work was supported in part by STARnet, a Semiconductor Research Corporation program sponsored by MARCO and DARPA, and in part by National Science Foundation under grant 1320808.}
}
\author{\IEEEauthorblockN{Shubham Jain, Ashish Ranjan, Kaushik Roy, Anand Raghunathan } \\
       \IEEEauthorblockA{School of Electrical and Computer Engineering, Purdue University \\ 
       \{jain130,aranjan,kaushik,raghunathan\}@purdue.edu \\[-3.5ex]}
}
\sloppy
\maketitle

\begin{abstract}
\noindent 
In-memory computing is a promising approach to addressing the processor-memory data transfer bottleneck in computing systems. We propose Spin-Transfer Torque Compute-in-Memory (STT-CiM), a design for in-memory computing with Spin-Transfer Torque Magnetic RAM (STT-MRAM). The unique properties of spintronic memory allow multiple wordlines within an array to be simultaneously enabled, opening up the possibility of directly sensing functions of the values stored in multiple rows using a single access. We propose modifications to STT-MRAM peripheral circuits that leverage this principle to perform logic, arithmetic and complex vector operations. We address the challenge of reliable in-memory computing under process variations by extending ECC schemes to detect and correct errors that occur during CiM operations. We also address the question of how STT-CiM should be integrated within a general-purpose computing system. To this end, we propose architectural enhancements to processor instruction sets and on-chip buses that enable STT-CiM to be utilized as a scratchpad memory. Finally, we present data mapping techniques to increase the effectiveness of STT-CiM. We evaluate STT-CiM using a device-to-architecture modeling framework, and integrate cycle-accurate models of STT-CiM with a commercial processor and on-chip bus (Nios II and Avalon from Intel). Our system-level evaluation shows that STT-CiM provides system-level performance improvements of 3.93x on average (upto 10.4x), and concurrently reduces memory system energy by 3.83x on average (upto 12.4x).

\end{abstract}



\vspace*{-0pt}
\section{Introduction}
\label{sec:introduction}
The growth in data processed and increase in the number of cores place high demands on the memory systems of modern computing platforms. Consequently, a growing fraction of transistors, area and power are utilized towards memories. CMOS memories (SRAM and embedded DRAM) have been the mainstays of memory design for the past several decades. However, recent technology scaling challenges in CMOS memories, along with an increased demand for memory capacity and performance, have fueled an active interest in alternative memory technologies.

Spintronic memories have emerged as a promising candidate for future memories due to several desirable attributes such as non-volatility, high density, and near-zero leakage. In particular, Spin Transfer Torque Magnetic RAM (STT-MRAM) has garnered significant interest with various prototype demonstrations and early commercial offerings~\cite{everspin,grandis,avalanche}. There have been several research efforts to boost the efficiency of STT-MRAM at the device, circuit and architectural levels~\cite{jog_dac,early_write,saibal_stt,yusung_stt,toshiba_1,partial_line,smullen_hpca,stt_llc,konWoo_yield_stt,umn_ecc,wangECC,STT_yield_wang,xfong_failure,mtj_20nm,approx_storage,3d_stt,lee_esttmram,stt_universal,hca_jadidi,multi_zhang,hmt_jishen,stt_trans_sizing,sttcache_siabal,coast,ikeda2008,wangRecofigSTTMRAM,dualpillar}. In this work, we explore {\em viz} in-memory computing with STT-MRAM. By exploiting the ability to simultaneously enable multiple wordlines within a memory array, we enhance STT-MRAM arrays to perform a range of arithmetic, logic and vector operations. We propose circuit and architectural techniques for reliable computation under process variations and to enable the proposed design to be used in a programmable processor based system.


In-memory computing is motivated by the observation that the movement of data from bit-cells in the memory to the processor and back (across the bit-lines, memory interface, and system interconnect) is a major performance and energy bottleneck in computing systems.
Efforts that have explored the closer integration of logic and memory are variedly referred to in the literature as logic-in-memory, computing-in-memory and processing-in-memory. These efforts may be classified into two categories -- moving logic closer to memory, or {\em near-memory computing}~\cite{iram,activepages,activedisks,diva,active_memory_cube,nmc_article,neurocube,nda,ndc,toppim,pim_enabled,interpolation_mem,hmc,hbm}, and performing computations within memory structures, or {\em in-memory computing}~\cite{tcam,compute_mem_sad,UMN_CRAM,pinatubo,MAGIC,memristiveLogic,bulk_bitwise2,acdimm,wangLiM,verma,reno,spindle,prime}, which is the focus of this work. In-memory computing reduces the number of memory accesses and the amount of data transferred between processor and memory, and exploits the wider internal bandwidth available within memory systems.

Our proposal is based on the observation that by enabling multiple wordlines simultaneously~\footnote{Note that this is much easier in STT-MRAM than in CMOS memories, due to the resistive nature of the bit-cells.} and sensing the effective resistance of each bit-line, it is possible to directly compute logic
functions of the values stored in the bit-cells. Based on this insight, we propose STT-CiM, a design for in-memory computing with STT-MRAM that can perform a range of arithmetic, logic, and vector operations. In STT-CiM, the core data array is the same as standard STT-MRAM; hence, memory density and the efficiency of read and write operations are maintained. Reliable sensing under the limited tunneling magneto-resistance (TMR) of STT-MRAM bit-cells is known to be a challenge~\cite{konWoo_yield_stt,umn_ecc,STT_yield_wang,xfong_failure,wangECC,wangRecofigSTTMRAM}, and we show that challenge this is further aggravated for in-memory computations. In order to enhance the robustness of STT-CiM under process variations, we extend error correction codes (ECC) to errors that occur during in-memory computations. To evaluate the benefits of STT-CiM, we utilize it as a scratchpad in the memory hierarchy of the Intel Nios II~\cite{nios2} processor. We propose enhancements to the on-chip bus and extend the instruction set of the processor to support compute-in-memory operations and expose them to software. We also present suitable data mapping techniques to maximize the benefits of STT-CiM.

We note that earlier efforts ({\em e.g.},~\cite{pinatubo}) have proposed enabling multiple wordlines to perform computations within Non-Volatile Memories (NVMs). Although our work shares this principle, we differ from previous work in several key aspects: (i) we address reliable in-memory computing under process variations, (ii) we go beyond bitwise logic operations to also perform arithmetic and vector operations, which are commonly present in modern computing workloads, and (iii) we propose architectural enhancements (bus and ISA extensions), and data mapping techniques to enable in-memory computation in the context of on-chip scratchpad memories.


In summary, the key contributions of this work are as follows:
\begin{itemize}
\item We explore compute-in-memory with spintronic memories as an approach to improving system performance and energy.
\item We propose STT-CiM, an enhanced STT-MRAM array that can perform a range of arithmetic, logic and vector compute-in-memory operations without modifying either the bit-cells or the core data array.  
\item We address a key challenge in STT-CiM, \emph{i.e.} reliably performing in-memory operations under process variation,  by demonstrating suitable error correction mechanisms. 
\item We propose extensions to the instruction set and on-chip bus to integrate STT-CiM into a programmable processor system and demonstrate the viability of these extensions using Intel's Nios II processor and Avalon on-chip bus.
\item  We evaluate the performance and energy benefits of STT-CiM, achieving average improvements of 3.83x (upto 12.4x) and 3.93x (upto 10.4x) in the total memory energy and system performance, respectively. 
\end{itemize}

The rest of the paper is organized as follows. Section~\ref{sec:related_work} presents an overview of prior research efforts related to in-memory computation. Section~\ref{sec:lim_background} provides the necessary background on STT-MRAM. Section~\ref{sec:lim_array} describes the STT-CiM design and how it supports in-memory computation. Section~\ref{sec:lim_architecture} outlines architectural enhancements for STT-CiM. Section~\ref{sec:exptsetup} describes the experimental methodology and experimental results are presented in section~\ref{sec:results}. Section~\ref{sec:conclusion} concludes the paper.

\vspace*{-0pt}

\vspace*{0pt}
\section{Related Work }
\label{sec:related_work}

\begin{figure}[htb]
  \vspace*{-4pt}
  \centering
  \includegraphics[width=\columnwidth]{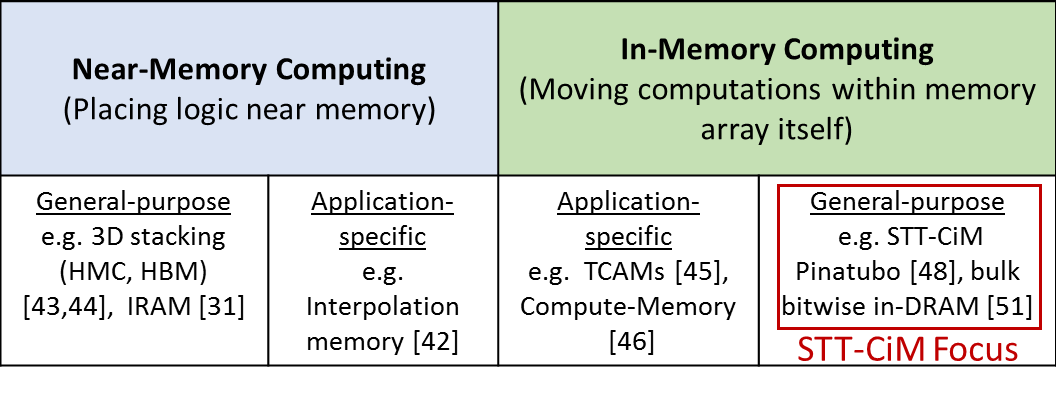}
  \vspace*{-12pt}
  \caption{ Related work: Overview}
  \label{fig:cim_nomenclature}
  \vspace*{-4pt}
\end{figure}

The closer integration of logic and memory is variedly referred to in the literature as logic-in-memory, computing-in-memory, and processing-in-memory. These efforts can be broadly classified into two categories, as shown in Figure~\ref{fig:cim_nomenclature}. We limit the scope of our discussion to approaches that improve the efficiency of active computation. For example, we do not discuss the embedding of non-volatile memory elements into a logic circuit~\cite{hanyu_lim_challenges,non_volatile_processor,matsunaga_prospects,meng_chang} in order to enable the system to shut-down and wakeup efficiently for improved power management.

Near-memory computing refers to bringing logic or processing units
closer to memory. Notwithstanding the closer integration, processing units
still remain distinct from memory arrays. Near-memory computing has been
explored at various levels of the memory
hierarchy~\cite{iram,activepages,activedisks,diva,active_memory_cube,neurocube,nda,ndc,toppim,pim_enabled,interpolation_mem,hmc,hbm}.
Intelligent RAM (IRAM)~\cite{iram} is an early example, which
integrated a processor and DRAM in the same chip to improve the
bandwidth between them. Embedding simple processing units within
each page of main memory~\cite{activepages} and within secondary
storage~\cite{activedisks} enables computations to be performed near
memory. An application-specific example of near-memory computation is
memory that can generate interpolated values, enabling the evaluation
of complex mathematical functions~\cite{interpolation_mem}. Near-memory
computing has gained significant interest in recent years, with industry
efforts like Hybrid Memory Cube (HMC)~\cite{hmc} and High Bandwidth Memory
(HBM)~\cite{hbm}. 

In-memory computing~\cite{tcam,compute_mem_sad,UMN_CRAM,pinatubo,bulk_bitwise2,acdimm,verma,wangLiM} integrates
logic operations into memory arrays, fundamentally blurring the distinction
between processing and memory. The key challenge of in-memory computing is to realize it without impacting the desirability of the resulting design as a standard memory
({\em i.e.}, density or efficiency of standard read and write
operations). Due to these constraints, in-memory computing is typically
limited to performing a small number of simple operations.

We can classify previous proposals for in-memory computing based on
whether they target application-specific or general-purpose
computations, and based on the underlying memory technology that they
consider. Application-specific examples of in-memory computing
include vector-matrix multiplication~\cite{verma,reno,spindle,prime} and sum-of-absolute difference~\cite{compute_mem_sad} computation. Ternary content-addressable
memory~\cite{tcam}, ROM-embedded RAM~\cite{rom_embedded_ram},
AC-DIMM~\cite{acdimm} and Micron's automata processor~\cite{micron_ap}
can also be viewed as examples of in-memory computing that target
specific operations such as pattern matching or evaluation of
transcendental functions.
Unlike these application-specific designs, we focus on embedding a
broader set of operations (arithmetic, logic and vector
operations) within memory.

In-memory evaluation of bitwise logic operations has been explored for memristive memories~\cite{pinatubo,memristiveLogic,MAGIC} and DRAM~\cite{bulk_bitwise2}.
Our work differs from these efforts in several important aspects. First, we focus on in-memory computing for spintronic memory, which involves fundamentally different prospects and design challenges. For example, the proposed operations are not destructive to the contents stored in the accessed bit-cells (unlike~\cite{bulk_bitwise2}). On the other hand, the much lower ratio of on to off resistance in spintronic memory leads to lower sensing margins. Second, we use a different sensing and reference generation circuitry, which enables us to natively realize a wider variety of operations. For example, the proposed design requires only one array access (unlike two in the case of~\cite{pinatubo}) to perform bit-wise XOR operations. 
Second, our design goes beyond bitwise logic operations and realizes
arithmetic as well as complex vector operations. Third, we propose architectural extensions (bus and ISA extensions) and data mapping techniques to enable in-memory computing within a general-purpose processor system. Finally, we address a key challenge associated with in-memory computing, \emph{viz.}, reliable operation under process variations. 

A different approach to in-memory computing with spintronic memories~\cite{UMN_CRAM} uses an extra transistor in each bit-cell (2T-1R cells), which sacrifices the density benefits of standard (1T-1R) STT-MRAM, while potentially enabling more complex functions to be evaluated within the array. In contrast, our proposal enables in-memory computation within a standard STT-MRAM array with no changes to the bit-cells. We note that a concurrent effort~\cite{wangLiM} has explored bit-wise AND/OR operations in STT-MRAM. The bit wise XOR operation cannot be realized atomically using the design proposed in~\cite{wangLiM}). Furthermore, these efforts restrict themselves to device and circuit level considerations, and do not address the architectural challenges of in-memory computing.

\vspace*{0pt}
\section{Background }
\label{sec:lim_background}
\begin{figure}[htb]
  \centering
  \vspace*{-0pt}
  \includegraphics[width=0.7\columnwidth]{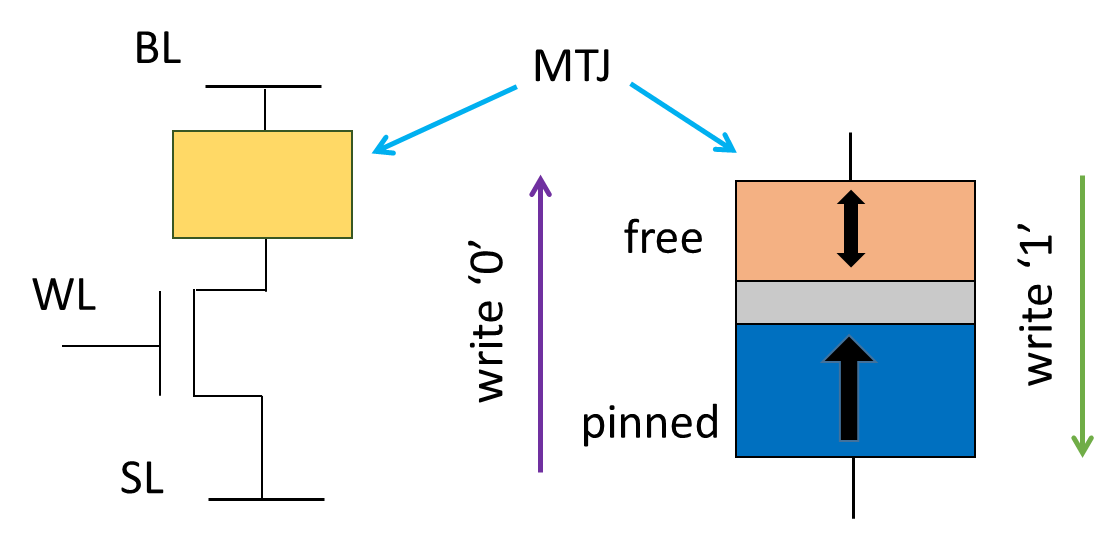}
  \vspace*{-0pt}
  \caption{ STT-MRAM bit-cell }
  \label{fig:stt_mram_bitcell}
  \vspace*{-0pt}
\end{figure}

\noindent An STT-MRAM bit-cell consists of an access transistor and a magnetic tunnel junction (MTJ), as shown in Figure~\ref{fig:stt_mram_bitcell}. An MTJ in turn consists of a pinned layer that has a fixed magnetic orientation and a free layer whose
magnetic orientation can be switched. The magnetic layers are separated by a tunnelling oxide. The relative magnetic orientation of the free and pinned layers determines the resistance offered by the MTJ (the resistance for the parallel configuration, $R_{P}$, is lower than the anti-parallel resistance, $R_{AP}$). The two resistance states encode a bit (we assume that parallel represents logic ``1", and anti-parallel represents logic ``0"). A read operation is performed by applying a bias ($V_{read}$) between the bitline (BL) and the source line (SL), and enabling the wordline (WL). The resultant current flowing through the bit-cell ($I_{P}$ or $I_{AP}$) is compared against a global reference to determine the logic state stored in the bit-cell.

A write is performed by passing a current greater than the critical switching current of the MTJ through the bit-cell. The logic value written is dependent on the direction of the write current as shown in Figure~\ref{fig:stt_mram_bitcell}. The write operation in STT-MRAM is stochastic in nature, and the duration and magnitude of the write current determines the write failure rate. Apart from write failures, STT-MRAMs are also subject to read decision failures, where the value stored in a bit-cell is incorrectly sensed due to process variations, and read disturb failures where a read operation inadvertently ends up writing into the bit-cell. These failures are addressed through a range of techniques including device and circuit optimization, manufacturing test and self-repair, and error correcting codes~\cite{konWoo_yield_stt,umn_ecc,STT_yield_wang,xfong_failure}. Apart from write/read failures, memories may also have failures due to thermal noise, which causes stochastic flipping in the bit-cells. However, such failures are negligible in STT-MRAM due to the high energy barrier between the two resistance states.

\vspace*{0pt}
\section{STT-MRAM based Compute-in-Memory (STT-CiM)}
\label{sec:lim_array}
\noindent In this section, we describe
STT-MRAM based Compute-in-Memory (STT-CiM), a design for in-memory
computing using standard STT-MRAM arrays.

\subsection{STT-CiM overview}
\label{subsec:cimconcept}

\begin{figure}[htb]
  \vspace*{-6pt}
  \centering
  \includegraphics[width=\columnwidth]{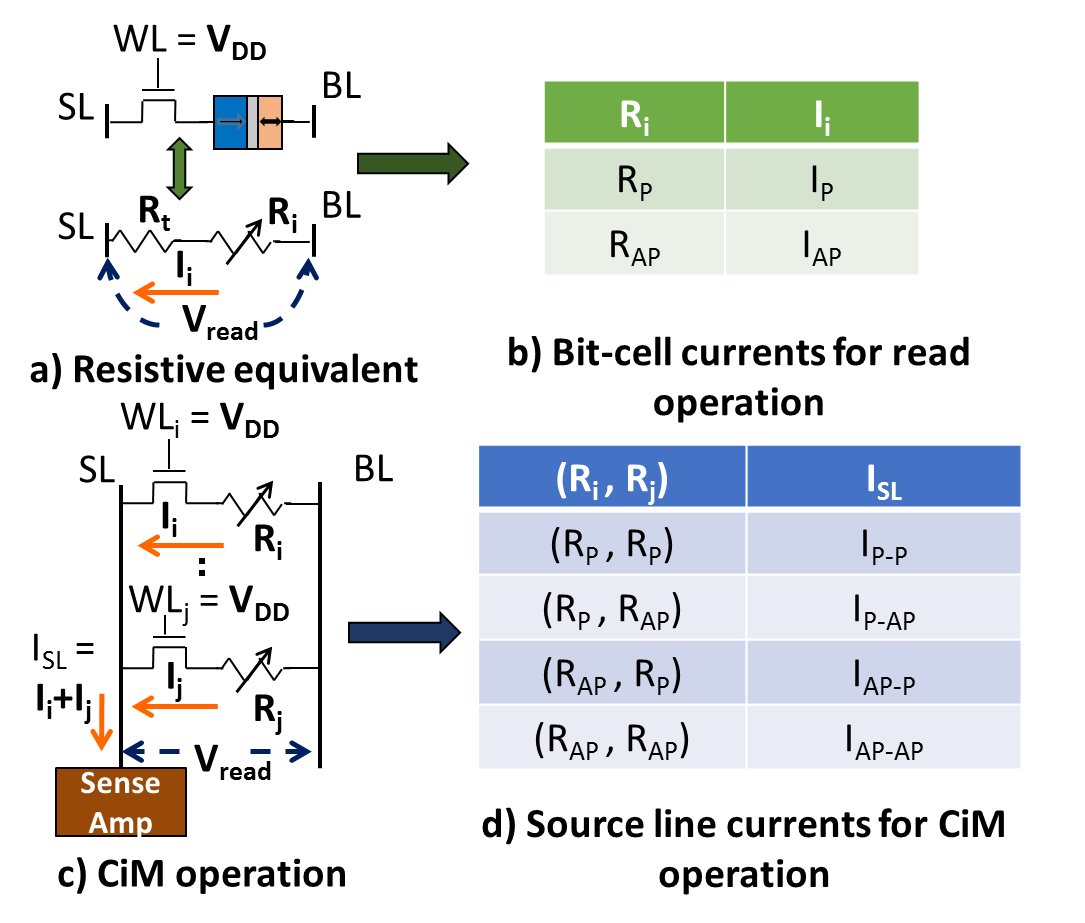}
  \vspace*{-10pt}
  \caption{ STT-CiM: Principle of operation}
  \label{fig:cim_operation_concept}
  \vspace*{-6pt}
\end{figure}

The key idea behind STT-CiM is to enable multiple wordlines simultaneously in an STT-MRAM array, leading to multiple bit-cells being connected to each bitline. With enhancements that we propose to the sensing and reference generation circuitry, we can directly compute logic functions of the enabled words. Note that such an operation is feasible in STT-MRAMs since the bit-cells are resistive, and since the write currents are typically much higher than read currents. In contrast, enabling multiple wordlines in SRAM can lead to short-circuit paths through the memory array, leading to loss of data stored in the bit-cells.

Figure~\ref{fig:cim_operation_concept} explains the principle of operation of STT-CiM. First, consider the resistive equivalent circuit of a single STT-MRAM bit-cell
shown in Figure~\ref{fig:cim_operation_concept}(a). $R_{t}$ represents the on-resistance of the access transistor and $R_i$ the resistance of the MTJ. When a voltage $V_{read}$ is applied between the bitline (BL) and the source line (SL), the net current $I_{i}$ flowing through the bit-cell can take two possible values depending on the MTJ configuration, as shown in Figure~\ref{fig:cim_operation_concept}(b). A read operation involves using a sensing mechanism to distinguish between these two current values.


Figure~\ref{fig:cim_operation_concept}(c) demonstrates a Compute-in-Memory (CiM) operation, where two wordlines ($WL_{i}$ and $WL_{j}$) are enabled, and a voltage bias ($V_{read}$) is applied to the bitline. The resultant current flowing through the SL (denoted $I_{SL}$) is a summation of the currents flowing through each of the bit-cells ($I_{i}$ and $I_{j}$), which in turn depends on the logic states stored in these bit-cells. The possible values of $I_{SL}$ are shown in Figure~\ref{fig:cim_operation_concept}(d). We propose enhanced sensing mechanisms to distinguish between these values and thereby compute logic functions of the values stored in the enabled bit-cells. We discuss the details of these operations in turn below.

\begin{figure}[htb]
  \vspace*{-6pt}
  \centering
  \includegraphics[width=\columnwidth]{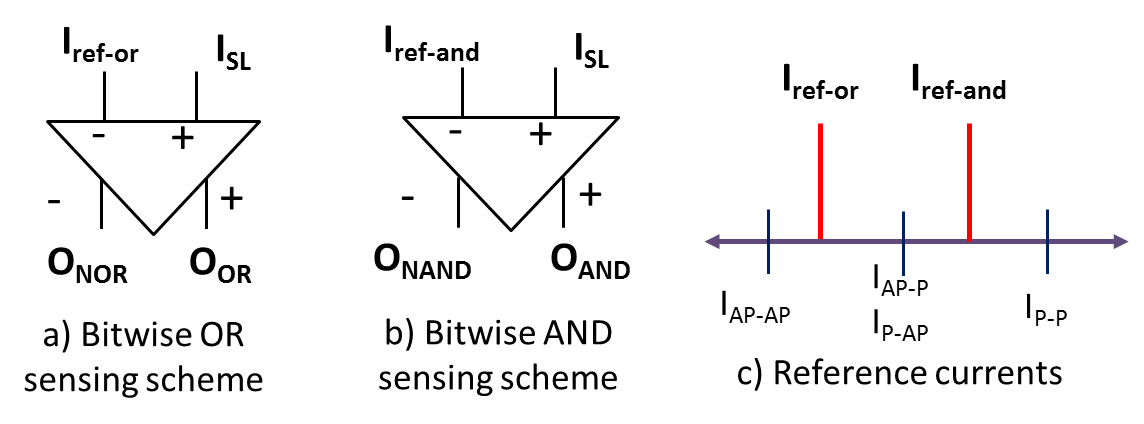}
  \vspace*{-10pt}
  \caption{ STT-CiM sensing schemes}
  \label{fig:cim_sensing}
  \vspace*{-4pt}
\end{figure}

{\bf\noindent Bitwise OR (NOR).} In order to realize logic OR and NOR operations, we use the sensing scheme shown in Figure~\ref{fig:cim_sensing}(a), where $I_{SL}$ is connected to the positive input of the sense amplifier and a reference current $I_{ref-or}$ is fed to its negative input. We choose $I_{ref-or}$ to be between $I_{AP-AP}$ and $I_{AP-P}$, as shown in Figure~\ref{fig:cim_sensing}(c). As a result, among the possible values of $I_{SL}$ [Figure~\ref{fig:cim_operation_concept}(d)], only $I_{SL}$ = $I_{AP-AP}$ is less than $I_{ref-or}$. Consequently, only the case where both bit-cells are in the AP configuration, {\em i.e.}, both store ``0", leads to an output of logic ``0" (``1") at the positive (negative) output of the sense amplifier, while all other cases lead to logic ``1" (``0"). Thus, the positive and negative outputs of the sense amplifier evaluate the logic OR and NOR of the values stored in the enabled bit-cells.

{\bf\noindent Bitwise AND (NAND).} A bitwise AND (NAND) operation is realized at the positive (negative) terminal of the sense amplifier by using the sensing scheme shown in Figure~\ref{fig:cim_sensing}(b). Note that in this scheme, a different  reference current ($I_{ref-and}$) is fed to the sense amplifier.  

{\bf\noindent Bitwise XOR.} A bitwise XOR operation is realized when the two sensing schemes shown in Figure~\ref{fig:cim_sensing} are used in tandem, and $O_{AND}$ and $O_{NOR}$ are fed to a CMOS NOR gate. In other words, $O_{XOR}$ = $O_{AND}$ NOR $O_{NOR}$. 

Table~\ref{tab:cim_sensing} summarizes the logic operations achieved using the two sensing schemes discussed above. Note that, all the logic operations described above are symmetric in nature, and hence it is not necessary to distinguish between the cases where the two bit-cells connected to a bitline store ``10" and ``01".

\vspace*{-0pt}
\newcolumntype{M}[1]{>{\centering\arraybackslash}m{#1}}
\renewcommand{\arraystretch}{1}
\renewcommand{\tabcolsep}{5pt}
\taburulecolor{black}
\begin{table}[hbtp]
  \vspace*{-2pt}
  \centering
  \caption{Possible outputs of various sensing schemes}
  \vspace*{-0pt}
  \begin{tabular}{ |M{13mm}|M{9mm}|M{9mm}|M{9mm}|M{9mm}|M{13mm}|}
	\hline
$I_{SL}$  &$O_{OR}$ &$O_{NOR}$ &$O_{AND}$ &$O_{NAND}$ &$O_{XOR}$ \\ \hline 
$I_{AP-AP}$ &0 &1 &0 &1 &0 \\ \hline
$I_{AP-P}$ &1&0 &0 &1 &1 \\ \hline
$I_{P-AP}$  &1 &0 &0 &1 &1  \\ \hline
$I_{P-P}$   &1 &0 &1 &0 &0  \\ \hline
   \end{tabular}    
  \vspace*{-0in}
  \label{tab:cim_sensing}
\end{table}
\vspace*{-4pt}
 
\begin{figure}[htb]
  \vspace*{-6pt}
  \centering
  \includegraphics[width=\columnwidth]{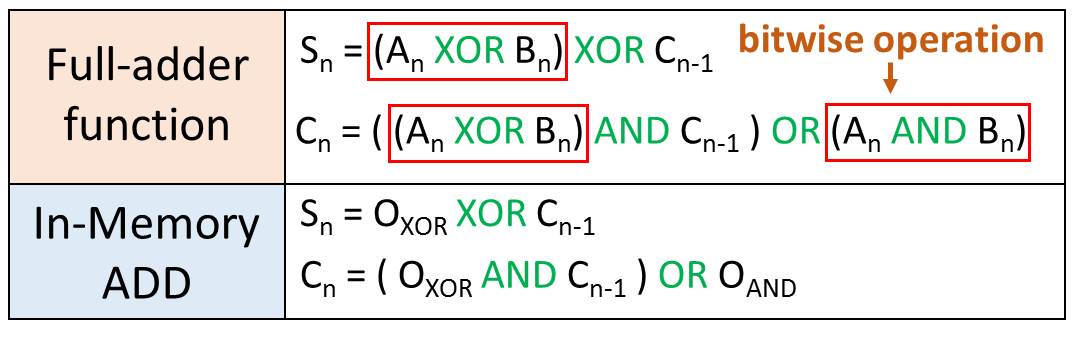}
  \vspace*{-10pt}
  \caption{ In-Memory ADD operation}
  \label{fig:add_eq}
  \vspace*{-10pt}
\end{figure}

{\bf\noindent ADD Operation.} An ADD operation is realized by leveraging the ability to concurrently perform multiple bitwise logical operations, as illustrated in Figure~\ref{fig:add_eq}. Suppose $A_n$ and $B_n$ (the n-th bits of two words, $A$ and $B$) are stored in two different bit-cells of the same column within an STT-CiM array. Suppose that we wish to compute the full-adder logic function (n-th stage of an adder that adds words $A$ and $B$). As shown in Figure~\ref{fig:add_eq}, $S_n$ (the sum) and $C_n$ (the carry out) can be computed using $A_n$ XOR $B_n$ and $A_n$ AND $B_n$, in addition to $C_{n-1}$ (carry input from the previous stage). 
Figure~\ref{fig:add_eq} also expresses the ADD operation in terms of the outputs of bitwise operations, $O_{AND}$ and $O_{XOR}$. Three additional logic gates are required to enable this computation. Note that the sensing schemes discussed enable us to perform the bitwise XOR and AND operations simultaneously, thereby performing an ADD operation with a {\em single array access}.

\begin{figure*}[htb]
  \vspace*{-8pt}
  \centering
  \includegraphics[width=\textwidth]{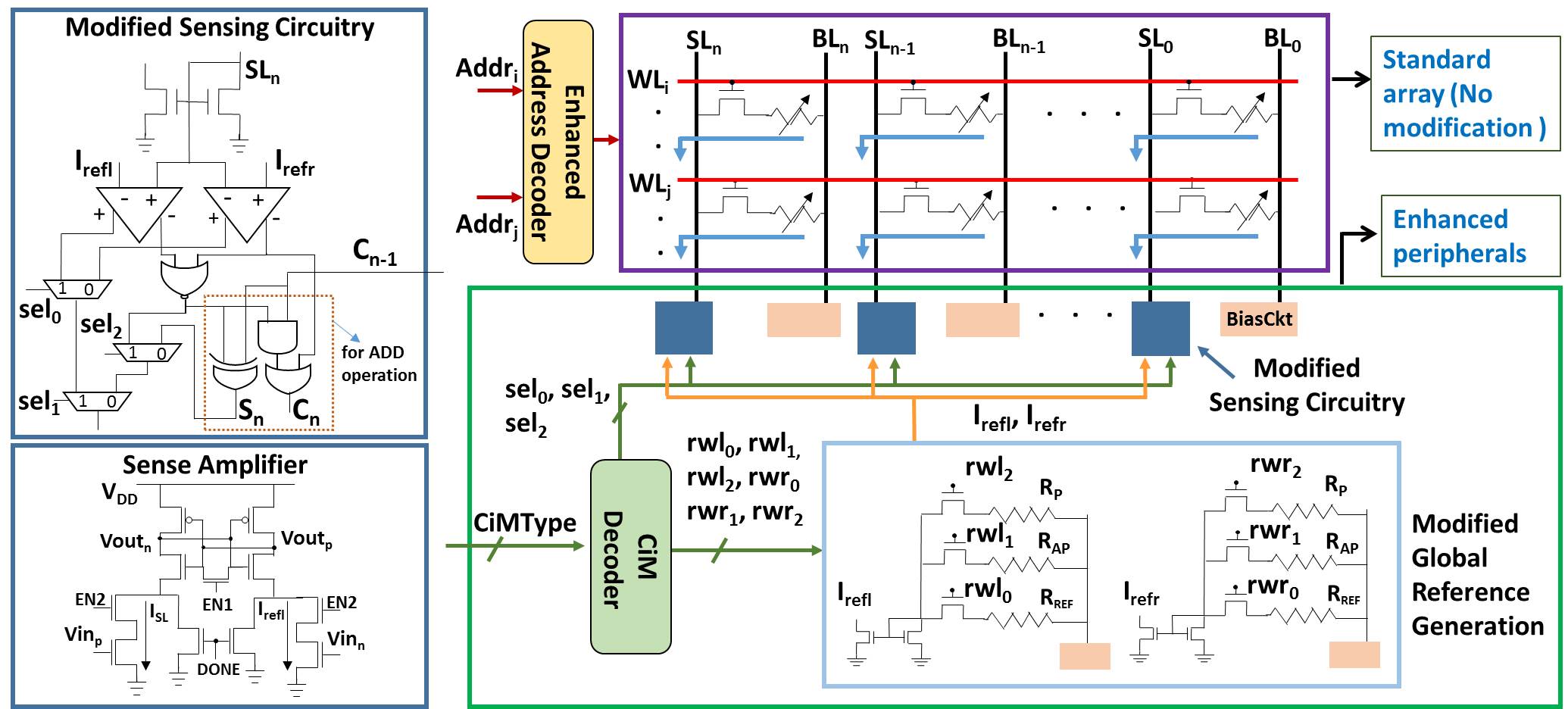}
  \vspace*{-10pt}
  \caption{ STT-CiM array structure }
  \label{fig:cimarrayfig}
  \vspace*{-8pt}
\end{figure*}

\subsection{STT-CiM array}
\label{subsec:array_structure}

In this section, we present the array-level design of STT-CiM using the circuit-level techniques described above. As shown in Figure~\ref{fig:cimarrayfig}, the proposed STT-CiM memory array takes an additional input CiMType that indicates the type of compute-in-memory operation that needs to be performed for every memory access. The CiM decoder interprets this input and generates appropriate control signals to perform the desired logic operation. In order to enable compute-in-memory operations, the read peripheral circuits present in each column (sensing circuit and global reference generation circuit in Figure~\ref{fig:cimarrayfig}) are enhanced, while the core data array remains the same as in standard STT-MRAM. The address (row) decoder needs to enable multiple wordlines for CiM operations. Specifically, we utilize two address decoders, with each decoding the corresponding input address. The corresponding outputs of the decoders are OR-ed and connected to each wordline. This configuration allows any of the two decoders to activate random WL locations. While the row decoder overhead is roughly doubled, it represents a small fraction of total area and power for configurations involving large arrays (1.8\% in our evaluation). The write peripheral circuits are unchanged, as write operations are identical to standard STT-MRAM. We next describe enhancements to sensing and reference generation circuits  to enable CiM operations.

{\bf\noindent Sensing circuitry.} Figure~\ref{fig:cimarrayfig} shows the sensing circuit enhanced to support all the logic operations discussed in Section~\ref{subsec:cimconcept}. It consists of two sense amplifiers, a CMOS NOR gate, three multiplexers and three additional logic gates for the ADD operation. We note that the area and power overheads associated with these enhancements are minimal since the sensing circuit constitutes a small fraction of the total memory area/power. As shown in the figure, the reference currents ($I_{refl}$, $I_{refr}$) produced by the global reference generation circuit are fed to the two sense amplifiers in order to realize the sensing schemes discussed in Section~\ref{subsec:cimconcept}. 
The three MUX control signals ($sel_{0}$, $sel_{1}$, $sel_{2}$) are generated by the CiM decoder to select the desired compute-in-memory operation.

\vspace*{-5pt}
\newcolumntype{M}[1]{>{\centering\arraybackslash}m{#1}}
\renewcommand{\arraystretch}{1.1}
\renewcommand{\tabcolsep}{5pt}
\taburulecolor{black}
\begin{table}[hbtp]
  \vspace*{-0pt}
  \centering
  \caption{STT-CIM operations control signals}
  \vspace*{-0pt}
  \begin{tabular}{ |M{10.0mm}|M{5.0mm}|M{5.0mm}|M{5.0mm}|M{5.0mm}|M{5.0mm}|M{5.0mm}|M{3.8mm}|M{3.8mm}|M{3.8mm}|}
	\hline
Operation  &$rwl_{0}$ &$rwl_{1}$ &$rwl_{2}$ &$rwr_{0}$ &$rwr_{1}$ &$rwr_{2}$ &$sel_{0}$ &$sel_{1}$ &$sel_{2}$\\ \hline \hline 
READ &1 &0 &0 &0 &0 &0 &1 &1 &x\\ \hline
NOT  &0 &0 &0 &1 &0 &0 &0 &1 &x\\ \hline
AND  &1 &0 &1 &0 &0 &0 &1 &1 &x\\ \hline
OR   &1 &1 &0 &0 &0 &0 &1 &1 &x\\ \hline
NAND &0 &0 &0 &1 &0 &1 &0 &1 &x\\ \hline
NOR  &0 &0 &0 &1 &1 &0 &0 &1 &x\\ \hline
XOR  &1 &1 &0 &1 &0 &1 &0 &0 &1\\ \hline
ADD  &1 &1 &0 &1 &0 &1 &0 &0 &0\\ \hline
    \end{tabular}    
  \vspace*{-0in}
  \label{tab:cim_control}
\end{table}
\vspace*{-5pt}

{\bf\noindent Reference generation.} Figure~\ref{fig:cimarrayfig} illustrates the modified reference generation circuit used to produce the additional reference currents necessary for the proposed sensing schemes. It includes two reference stacks, one for each of the two sense amplifiers in the sensing circuit. Each stack consists of three bit-cells programmed to offer resistances $R_{P}$, $R_{AP}$ and $R_{REF}$, respectively. $R_{REF}$~\footnote{$R_{AP}$ $>$ $R_{REF}$ $>$ $R_{P}$} represents the fixed resistance reference MTJ used in a standard STT-MRAM to perform read operations. 
The CiM decoder generates control signals ($rwl_{0}$, $rwl_{1}$, .... $rwr_{1}$, $rwr_{2}$) that enable a subset of these bit-cells in the reference stacks, which in turn produces the desired reference currents. Table~\ref{tab:cim_control} presents the values of these control signals so as to achieve the required reference currents. 

The STT-CiM array can perform both regular memory operations and a range of CiM operations. The normal read operation is performed by enabling a single wordline and setting $sel_{0}$, $sel_{1}$, and $rwl_{0}$ to logic '1'. On the other hand, a CiM operation is performed by enabling two wordlines and setting CiMType to the appropriate value, which results in computing the desired function of the enabled words. The control signal values for a read operation as well as CiM operations are shown in Table~\ref{tab:cim_control}. 

  

\subsection{CiM operation under process-variations}
\label{subsec:cim_op_read_margin}

\begin{figure}[htb]
  \vspace*{-10pt}
  \centering
  \includegraphics[width=\columnwidth]{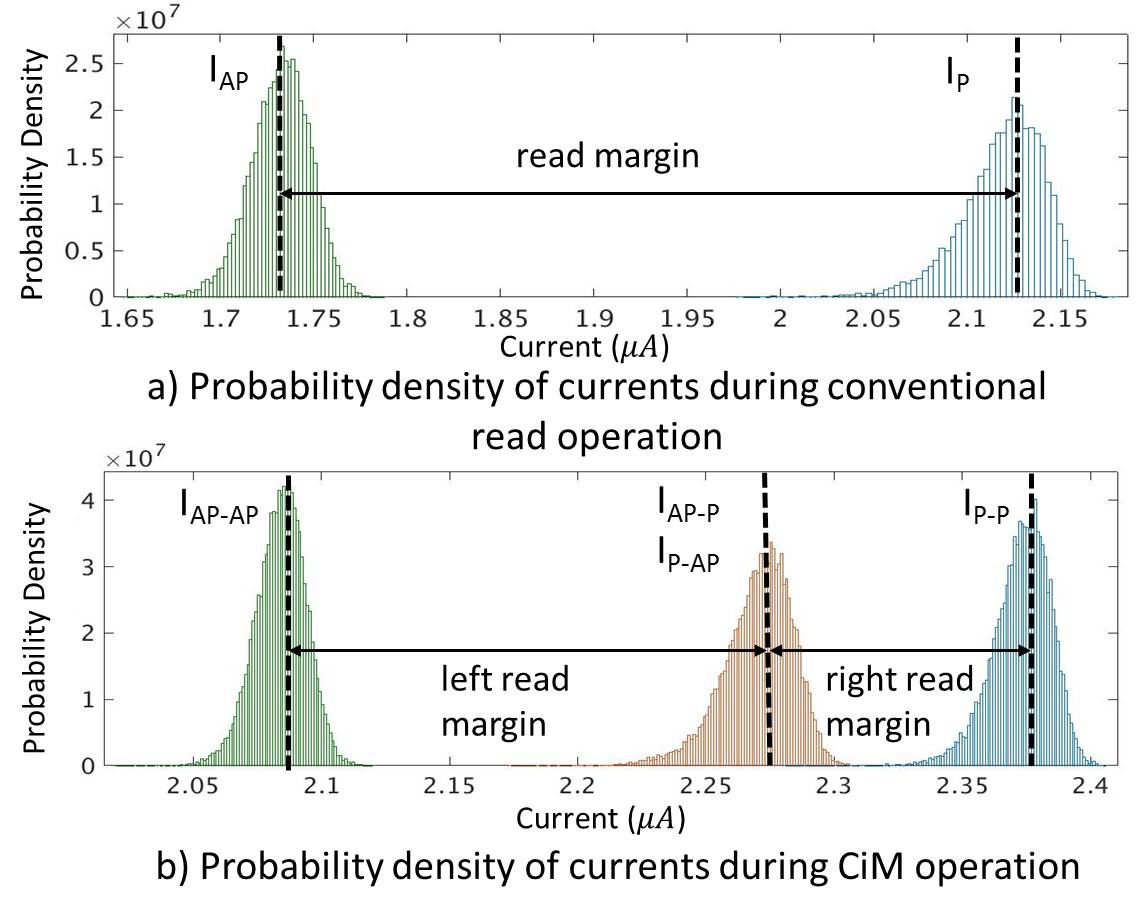}
  \vspace*{-14pt}
  \caption{ Probability density distribution of $I_{SL}$ under process variations during read and CiM operations }
  \label{fig:varresult}
  \vspace*{-10pt}
\end{figure}

The STT-CiM array suffers from the same failure mechanisms (read disturb failures, read decision failures and write failures) that are observed in standard STT-MRAM. In this section, we compare the failure rates in STT-CiM and standard STT-MRAM. Normal read/write operations in STT-CiM have the same failure rate as in a standard STT-MRAM, since the read/write mechanisms are identical. However, CiM operations differ in their failure rates, since the currents that flow through each bit-cell differ when enabling two wordlines simultaneously. In order to analyze the read disturb and read decision failures under process variations for CiM operations, we performed a Monte-carlo circuit-level simulation on 1 million samples considering variations in MTJ oxide thickness ($\sigma/\mu$ = 2 \%), transistor $V_{T}$ ($\sigma/\mu$ = 5\%), and MTJ cross sectional area ($\sigma/\mu$ = 5\%)~\cite{konWoo_yield_stt}. Figure~\ref{fig:varresult} shows the probability density distribution of the possible currents obtained during read and CiM operations on these 1 million samples.


{\bf\noindent CiM disturb failures.} As shown in Figure~\ref{fig:varresult}, the overall current flowing through the source line is slightly higher in case of a CiM operation as compared to a normal read. However this increased current is divided between the two parallel paths, and consequently the net read current flowing through each bit-cell (MTJ) is reduced. Hence, the read disturb failure rate is lower for CiM operations than for normal read operations.

{\bf\noindent CiM decision failures.} The net current flowing through the source line ($I_{SL}$) in case of a CiM operation can have 3 possible values, \emph{i.e.}, $I_{P-P}$, $I_{AP-P}$ ($I_{P-AP}$), $I_{AP-AP}$. A read decision failure occurs during a CiM operation when the current $I_{P-P}$ is interpreted as $I_{AP-P}$ (or vice versa), or when $I_{AP-AP}$ is inferred as $I_{AP-P}$ (or vice versa). In contrast to normal reads, CiM operations have two read margins --- one between $I_{P-P}$ and $I_{AP-P}$ and another between $I_{AP-P}$ and $I_{AP-AP}$ [Figure~\ref{fig:varresult}(b)]. Our simulation results show that the read margins for CiM operations are lower as compared to normal reads, therefore they are more prone to decision failures. Moreover, the read margins in CiM operations are unequal~\footnote{Although resistances may be equally separated, the currents are not since they depend inversely on resistance.}. Thus, we have more failures arising due to the read margin between $I_{P-P}$ and $I_{AP-P}$. 

\begin{figure}[htb]
  \vspace*{-8pt}
  \centering
  \includegraphics[width=\columnwidth]{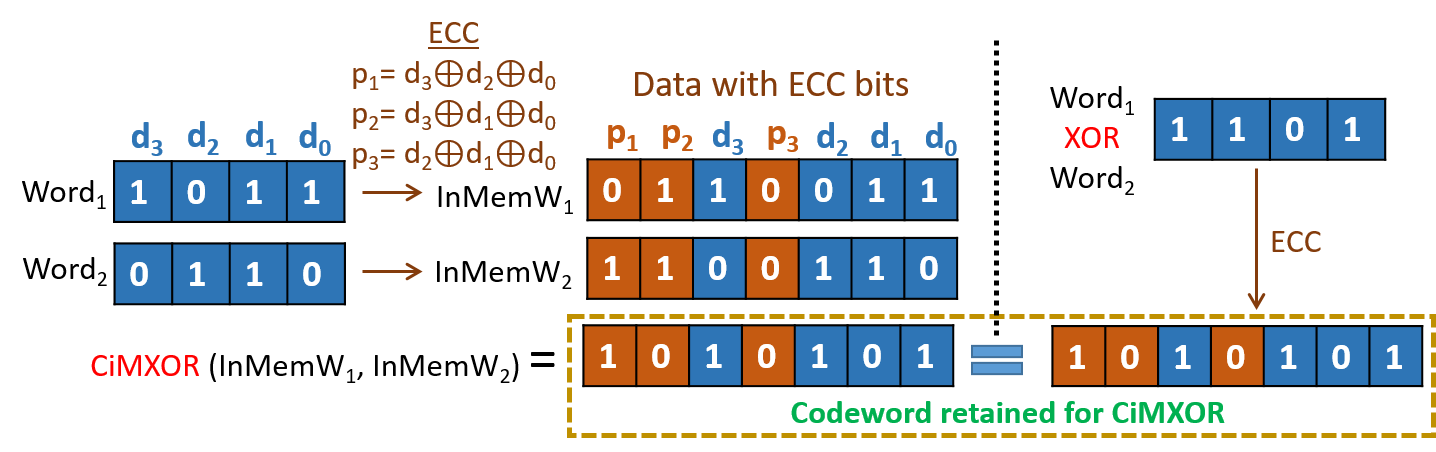}
  \vspace*{-8pt}
  \caption{ Codeword retention property of CIM XOR}
  \label{fig:codeword_CiMXOR}
  \vspace*{-8pt}
\end{figure}

{\bf\noindent ECC for STT-CiM.} In order to mitigate these failures in STT-MRAM, various ECC schemes have been previously explored~\cite{konWoo_yield_stt,umn_ecc,wangECC}. We show that ECC techniques that provide single error correction and double error detection (SECDED) and double error correction and triple error detection (DECTED) can be used to address the decision failures in CiM operations as well. This is feasible because the codeword properties for most ECC codes are retained for a CiM XOR operation. Figure~\ref{fig:codeword_CiMXOR} shows the codeword retention property of a CiM XOR operation using a simple Hamming code. As shown in the figure, $word_{1}$ and $word_{2}$ are augmented with ECC bits ($p_{1}$, $p_{2}$, $p_{3}$) and stored in memory as $InMemW_{1}$ and $InMemW_{2}$ respectively. A CiM XOR operation performed on these stored words ($InMemW_{1}$, $InMemW_{2}$) results in the ECC codeword for $word_{1}$ XOR $word_{2}$, therefore the codewords are preserved for CiM XORs. We leverage this codeword retention property of CiM XORs to detect and correct errors in \emph{all CiM operations}. This is enabled by the fact that STT-CiM always computes bitwise XOR (CiM XOR) irrespective of the CiM operation that is being performed. 

\begin{figure}[htb]
  \vspace*{-8pt}
  \centering
  \includegraphics[width=\columnwidth]{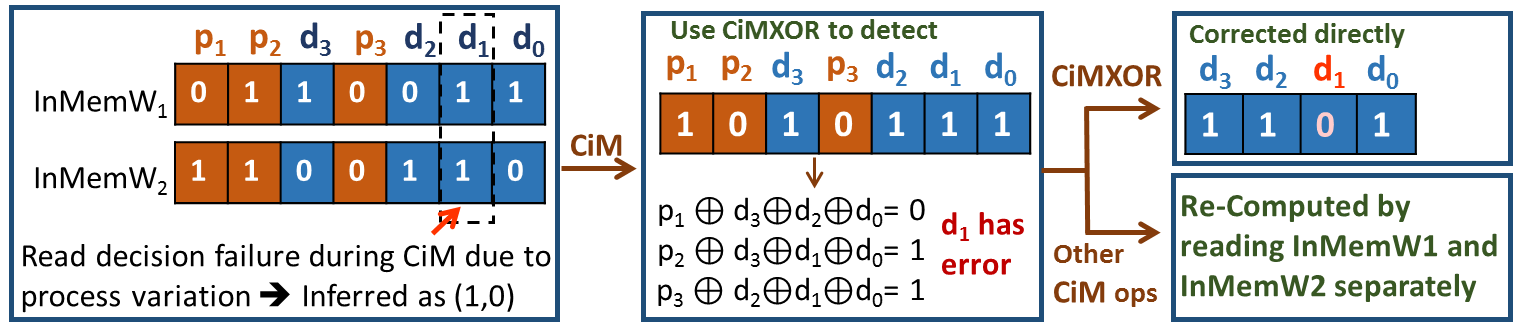}
  \vspace*{-8pt}
  \caption{Error detection and correction for CiM operations}
  \label{fig:ecc_CiMXOR}
  \vspace*{-8pt}
\end{figure}

We demonstrate the proposed error detection and correction mechanism for CiM operations in Figure~\ref{fig:ecc_CiMXOR}. 
Let us assume that data bit $d_{1}$ suffers from a decision failure during CiM operations, as shown in the figure. As a result, the combination of logic 1 and 1 in the two bit-cells ($I_{P-P}$) is inferred as logic 1 and 0 ($I_{AP-P}$), leading to erroneous CiM outputs. An error detection logic operating on the CiM XOR output (Figure~\ref{fig:ecc_CiMXOR}) detects an error in the $d_{1}$ data bit. This error can be corrected directly for a CiM XOR operation by simply flipping the erroneous bit. For other CiM operations, we perform two conventional reads on words $InMemW_{1}$ and $InMemW_{2}$, and correct the erroneous bits by recomputing them using an error detection and correction unit (discussed in section~\ref{sec:lim_architecture}). Note that such corrections lead to overheads, as we need to access memory array 3 times (compared to 2 times in STT-MRAM). However, our variation analysis shows that error corrections on CiM operations are infrequent, leading to overall improvements.   

{\bf\noindent ECC design methodology.}
We use the methodology employed in~\cite{konWoo_yield_stt} to determine ECC requirements for both the baseline STT-MRAM and the proposed STT-CiM design. The approach uses circuit level simulations to determine the bit-level error probability, which is then used to estimate the array level yield. Moreover, the ECC scheme is selected based on the target yield requirement. Our simulation shows that 1 bit failure probability of normal reads and CiM operations are 4.2x$10^{-8}$ and 
6x$10^{-5}$ respectively. With these obtained bit-level failure rates and assuming a target yield of 99$\%$, the ECC requirement for 1MB STT-MRAM is single error correction and double error detection (SECDED), whereas for 1MB STT-CiM is three error correction and four error detection (3EC4ED). Note that the overheads of the ECC schemes~\cite{BCH_Decoder,konWoo_yield_stt} are fully considered and reflected in our experimental results. Moreover, our simulation shows that the probability of CiM operations having errors is at most 0.1, \emph{i.e.}, no more than 1 in 10 CiM operations will have an error. Errors on all CiM operatons are detected by using 3EC4ED code on the XOR output. Detected errors are directly corrected for CiM XORs using the 3EC4ED code, 
and by reverting to near-memory computation for other CiM operations. 

Apart from ECC schemes, STT-CiM can also leverage various reliability improvement techniques proposed for STT-MRAMS~\cite{coast,ikeda2008,wangRecofigSTTMRAM,dualpillar}. Further, recent efforts~\cite{coast,ikeda2008} that increase the Tunneling Magneto Resistance (TMR) of the MTJ and improve sensing margins will reduce read failure in CiM operations as well. These techniques can be used along with ECC to cost-effectively mitigate failures in STT-CiM operations.

\vspace*{-0pt}
\section{STT-CiM Architecture}
\label{sec:lim_architecture}
\noindent In order to evaluate the application-level benefits of STT-CiM, we integrate it as a scratchpad memory within the memory hierarchy of a programmable processor~\cite{nios2}. This section describes architectural enhancements for STT-CiM and hardware/software optimizations to increase its efficiency.

\subsection{Optimizations for STT-CiM }
\label{subsec:vector_op}

In order to further the efficiency improvements obtained by STT-CiM, we propose additional optimizations.    

\begin{figure}[htb]
  \centering
  \vspace*{-8pt}
  \includegraphics[width=\columnwidth]{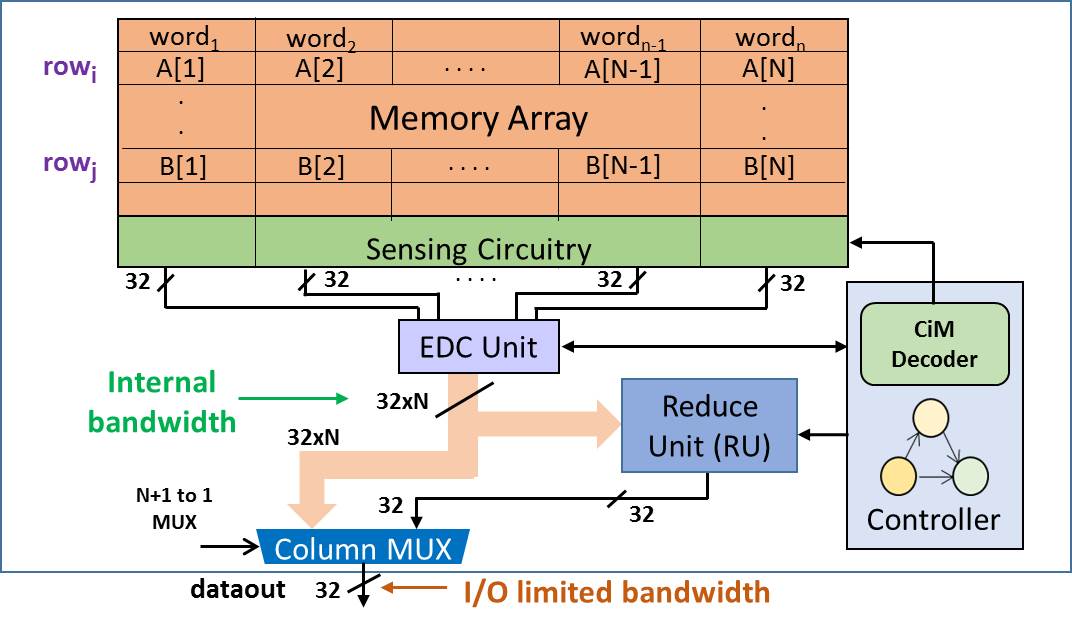}
  \vspace*{-12pt}
  \caption{ STT-CiM supporting In-memory vector operation  }
  \label{fig:cimvector_op}
  \vspace*{-6pt}
\end{figure}

{\bf\noindent Vector CiM operations.} Modern computing workloads exhibit significant data parallelism. To further enhance the efficiency of STT-CiM for data-parallel computations, we introduce Vector Compute-in-Memory (VCiM) operations. The key idea behind VCiM operations is to perform CiM operations on all the elements of a vector concurrently. Figure~\ref{fig:cimvector_op} shows how the internal memory bandwidth (32xN bits) can be significantly larger than the limited I/O bandwidth (32 bits) visible to the processor. We exploit the memory's internal bandwidth to perform vector operations (N words wide) within STT-CiM. 

Note that the data resulting from a vector operation may also be a vector, and hence transferring it back to the processor is subject to the limited I/O bandwidth. To address this issue, we observe that vector operatons are often followed by reduction 
operations. For example, a vector dot-product involves element-wise multiplication of two vectors followed by a summation (reduction) of the resulting vector of products to produce a scalar value. Based on this observation, we introduce a Reduce Unit (RU) before the column multiplexer, as shown in Figure~\ref{fig:cimvector_op}. The RU takes an array of data elements as inputs and reduces it to a single data element. The RU can support various reduction operations such as summation, Euclidean distance, L1 and L2 norm, zero-comparision, {\em etc.}, of which two are described
in Table~\ref{tab:reduceop}. Consider the computation of $\sum_{i=1}^{N} A[i] + B[i]$, where arrays A and B are stored in rows i and j respectively (shown in Figure~\ref{fig:cimvector_op}). To compute the desired function using a VCiM operation, we activate rows i and j simultaneously, and configure the sensing circuitry to perform an ADD operation and the RU to perform accumulation of the resulting output. Note that the summation would require 2N memory accesses in a conventional memory. With scalar CiM operations, it would require N memory accesses. With the proposed VCiM operations, only a single memory access is required. 

The overheads of the RU depend on two factors: (i) the number of different reduction operations supported, and (ii) the maximum vector length allowed (can be between 2 to N words). To limit the overheads, we restrict our design to vector lengths of 4 and 8. 



{\bf\noindent Error Detection and Correction.} To enable correction of erroneous bits for CiM operations, we introduce an Error Detection and Correction (EDC) unit that implements the 3EC4ED ECC scheme. The EDC unit checks for errors using the CiM XOR output (recall that the XOR is evaluated along with all CiM operations) and signals the controller (shown in Figure~\ref{fig:cimvector_op}) upon detection of erroneous computations. Upon receiving this error detection signal, the controller performs the required corrective actions.

\vspace*{-0pt}
\newcolumntype{M}[1]{>{\centering\arraybackslash}m{#1}}
\renewcommand{\arraystretch}{1.1}
\renewcommand{\tabcolsep}{5pt}
\begin{table}[hbtp]
  \vspace*{-4pt}
  \centering
  \caption{Examples of reduction operations }
  \vspace*{-0pt}
  \begin{tabular}{ |M{20mm}|M{50mm}|}
	\hline
Type  &Function  \\ 
      & RuOut = f ($IN_{1}$, $IN_{2}$,...,$IN_{N}$) \\ \hline \hline 
Summation &RuOut=$IN_{1}$ + $IN_{2}$+ .. + $IN_{N}$ \\ \hline
Zero-Compare  &RuOut[k] = ($IN_{k}$ == 0) ? 0 : 1 \\ \hline
    \end{tabular}    
  \vspace*{-0in}
  \label{tab:reduceop}
\end{table}
\vspace*{-10pt}

\subsection{Architectural Extensions for STT-CiM }
\label{subsec:archi_isa}

To integrate STT-CiM in a programmable processor based system, we propose the following architectural enhancements.

{\bf\noindent ISA extension.} We extend the ISA of a programmable processor to support CiM operations. To this end, we introduce a set of new instructions in the ISA (CiMXOR, CiMNOT,CiMAND, CiMADD ...) that are used to invoke the different types of operations that can be performed in the STT-CiM array. In a load instruction, the requested address is sent to the memory, and the memory returns the data stored at the addressed location. However, in the case of a CiM instruction, the processor is required to provide addresses of two memory locations instead of a single one, and the memory operates on the two data values to return the final output.

\vspace*{-10pt}
\begin{equation}
   \begin{aligned}
   \begin{split}
 	& \text{Format:   }\textbf{ Opcode   }  \text{Reg1  } \text{Reg2  } \text{Reg3  } \\
	& \text{Example:  }\textbf{CiMXOR   }  \text{$R_{ADDR1}$  } \text{$R_{ADDR2}$ } \text{$R_{DEST}$ }
	\end{split}
	\end{aligned}
	\label{eq:cim_opcode_type}
\end{equation}
\vspace*{-5pt}

Equation~\ref{eq:cim_opcode_type} shows the format of a CiM instruction with an example. As shown, both the addresses required to perform CiMXOR operations are provided through registers. The format is similar to a regular arithmetic instruction that accesses two register values, performs the computation, and stores the result back in a register.

\begin{figure}[htb]
  \vspace*{-10pt}
  \includegraphics[width=\columnwidth]{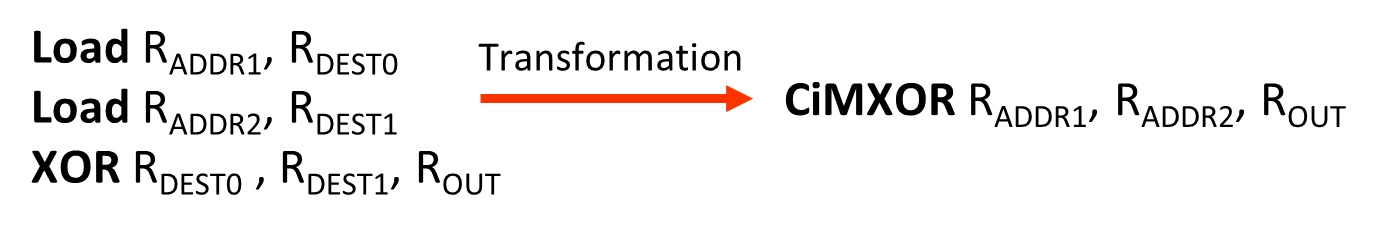}
  \vspace*{-16pt}
  \caption{ Program transformation for CiMXOR }
  \label{fig:cimopcode_trans}
  \vspace*{-8pt}
\end{figure}

{\bf\noindent Program transformation.} To exploit the proposed CiM instructions at the application-level, an assembly-level program transformation is performed, wherein specific sequences of instructions in the compiled program are mapped to suitable CiM instructions in the ISA. Figure~\ref{fig:cimopcode_trans} shows an example transformation where two load instructions followed by an XOR instruction are mapped to a single CiMXOR instruction. 

{\bf\noindent Bus and interface support.}  In a programmable processor based system, the processor and the memory communicate via a system bus or on-chip network~\footnote{While we consider the case of a shared bus for illustration, the same enhancements can be applied to more complex interconnect networks.}. This makes it essential to analyze the impact of CiM operations on the bus and the corresponding bus interface. As discussed above, a CiM operation is similar to a load instruction with the key difference that it sends two addresses to the memory. Conventional system buses only allow sending a single address onto the bus via the address channel. In order to send the second address for CiM operations, we utilize the unused writedata channel of the system bus, which is unutilized during a CiM operation. Besides the two addresses, the processor also sends the type of CiM operation (CIMType) that needs to be performed. Note that it may be possible to overlay the CIMType signal onto the existing bus control signals; however, such optimizations strongly depend on the specifics of the bus protocol being used. In our design, we assume that 3 control bits are added to the bus to carry CIMType, and account for the resulting overheads in our experiments.


\subsection{Data Mapping}

\label{subsec:dataplacement_isa}

In order to perform a CiM instruction, the locations of its operands in memory
must satisfy certain constraints. Let us consider a memory organization consisting of several banks where each bank is an array that contains rows and columns. In this case, a CiM operation can be performed on two data elements only if they satisfy three key criteria: (i) they are stored in the same bank,
(ii) they are mapped to different rows, and (iii) they are stored
in the same set of columns.

\begin{figure}[htb]
  \vspace*{-10pt}
  \includegraphics[width=\columnwidth]{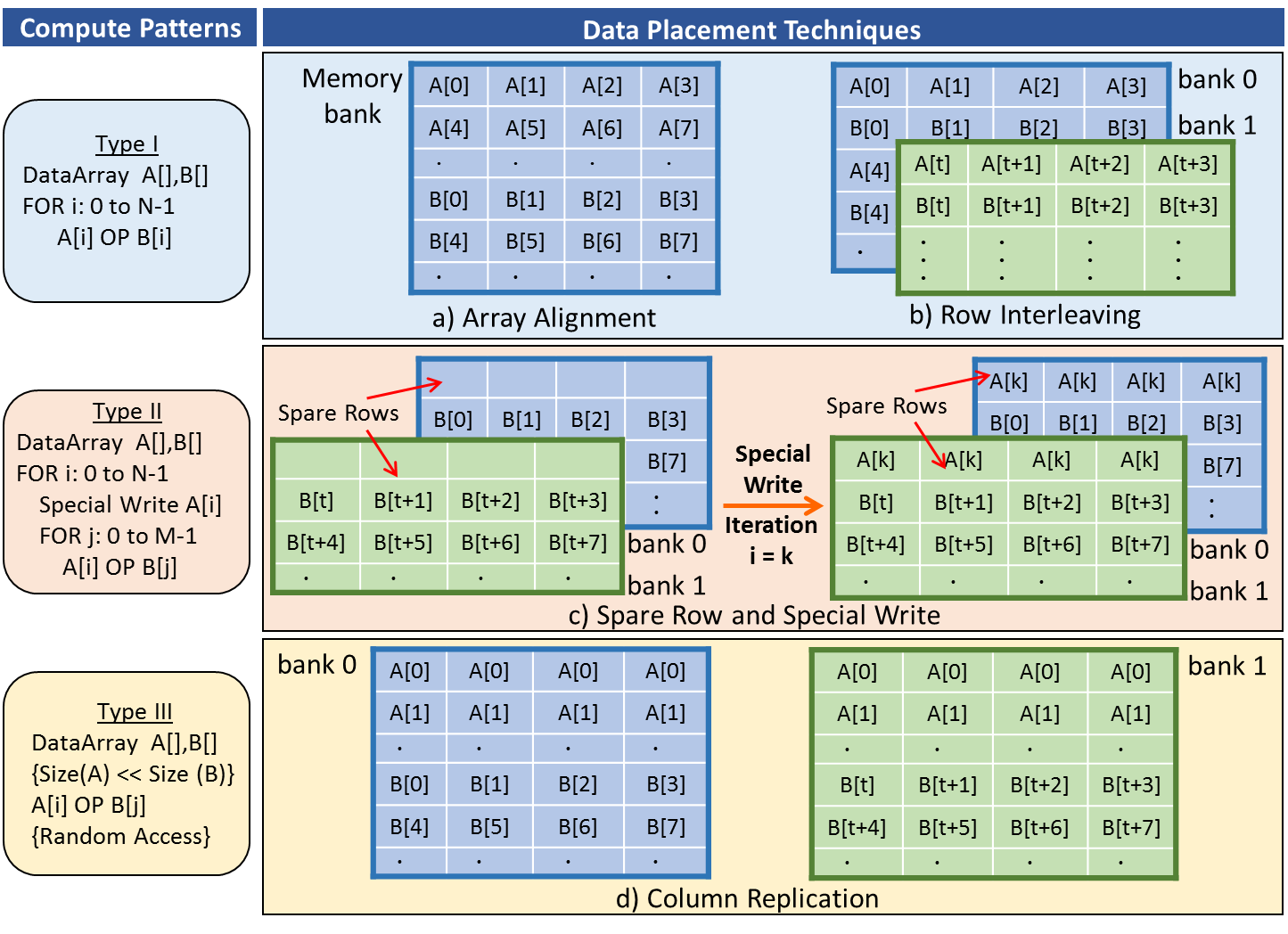}
  \vspace*{-8pt}
  \caption{Data mapping for various computation patterns}
  \label{fig:cim_placement}
  \vspace*{-6pt}
\end{figure}

Consequently, a suitable data placement technique is required that maximizes the use of CiM operations. We observe that the target applications for STT-CiM have well-defined computation patterns, facilitating such a data placement. Figure~\ref{fig:cim_placement} shows three general computation patterns. We next discuss these compute patterns and the corresponding data placement techniques.

{\bf\noindent Type I.} This pattern, shown in the top row of Figure~\ref{fig:cim_placement}, involves element-to-element operations (OPs) between two arrays, \emph{e.g.}, A and B. In order to effectively utilize STT-CiM for this compute pattern, we utilize the \emph{array alignment} technique (shown in Figure~\ref{fig:cim_placement}(a)) that ensures alignment of elements A[i] and B[i] of arrays A and B for any value of i. This enables the conversion of operation A[i] OP B[i] into a CiM operation. An extension to this technique is the \emph{row-interleaved placement} shown in Figure~\ref{fig:cim_placement}(b). This technique is applicable to larger data structures that do not fully reside in same memory bank. It ensures that the corresponding elements, \emph{i.e.}, A[i] and B[i], are mapped to the same bank for any value of i, and satisfy the alignment criteria for a CiM operation.

{\bf\noindent Type II.}  This pattern, shown in the middle row of Figure~\ref{fig:cim_placement}, involves a nested loop in which the inner loop iteration consists of a single element of array A being operated with several elements of array B. For this one-to-many compute pattern, we introduce a \emph{spare row} technique for data alignment. In this technique, a spare row is reserved in each memory bank to store
copies of an element of A. As shown in Figure~\ref{fig:cim_placement}(c), in the $k^{th}$ iteration of the outer for-loop, a special write
operation is used to fill the spare rows in all banks with A[k]. This results in each element of array B becoming aligned with a copy of A[k], thereby allowing CiM operations to be performed on them. Note that the special write operation introduces energy and performance overheads, but this overhead is amortized over all inner loop iterations, and is observed to be quite insignificant in our evaluations.

{\bf\noindent Type III.} In this pattern, shown in the bottom row of Figure~\ref{fig:cim_placement}, operations are performed on an element drawn from a small array A and an element from a much larger array B. The elements are selected arbitrarily, {\em i.e.}, without any predictable pattern. For example, consider when a small sequence of characters needs to be searched within a much larger input string. For this pattern, we propose a \emph{column replication} technique to enable CiM operations, as shown in Figure~\ref{fig:cim_placement}(d). In this technique, a single element of the small array A is replicated across columns to fill an entire row. This ensures that each element of A is aligned with every element of B, enabling a CiM operation to be utilized. Note that the initial overhead due to data replication is very small, as it pales in comparison to the number of memory accesses to the larger array.

\vspace*{-0pt}
\section{Experimental Methodology}
\label{sec:exptsetup}
\begin{figure}[htb]
  \vspace*{-6pt}
  \centering
  \includegraphics[width=0.9\columnwidth]{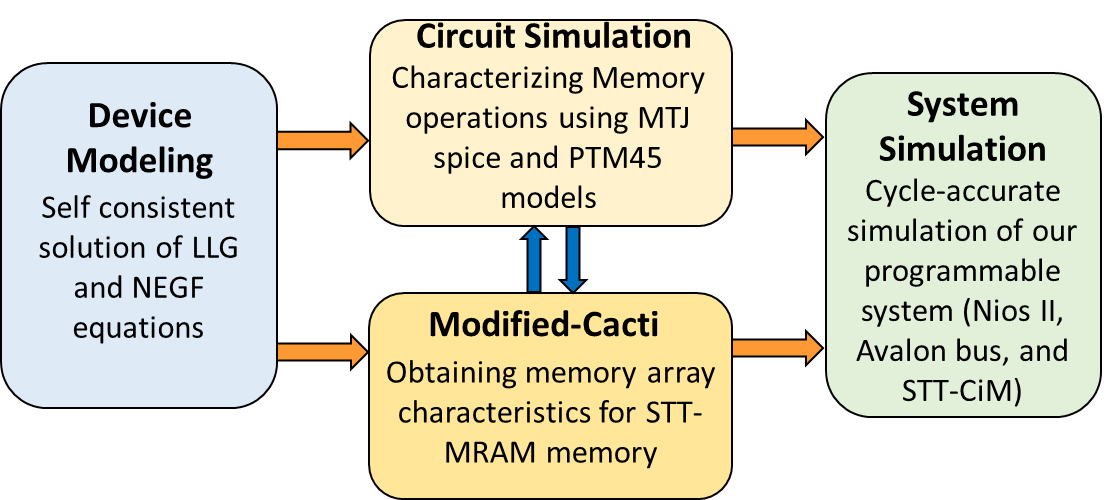}
  \vspace*{-2pt}
  \caption{STT-CiM device-to-architecture evaluation framework}
  \label{fig:cim_framework}
  \vspace*{-5pt}
\end{figure}

{\bf \noindent}  In this section, we discuss the device-to-architecture simulation framework (Figure~\ref{fig:cim_framework}) and application benchmarks used to evaluate the performance and energy benefits of STT-CiM at the array-level and system-level.

\begin{figure}[htb]
  \vspace*{-6pt}
  \centering
  \includegraphics[width=\columnwidth]{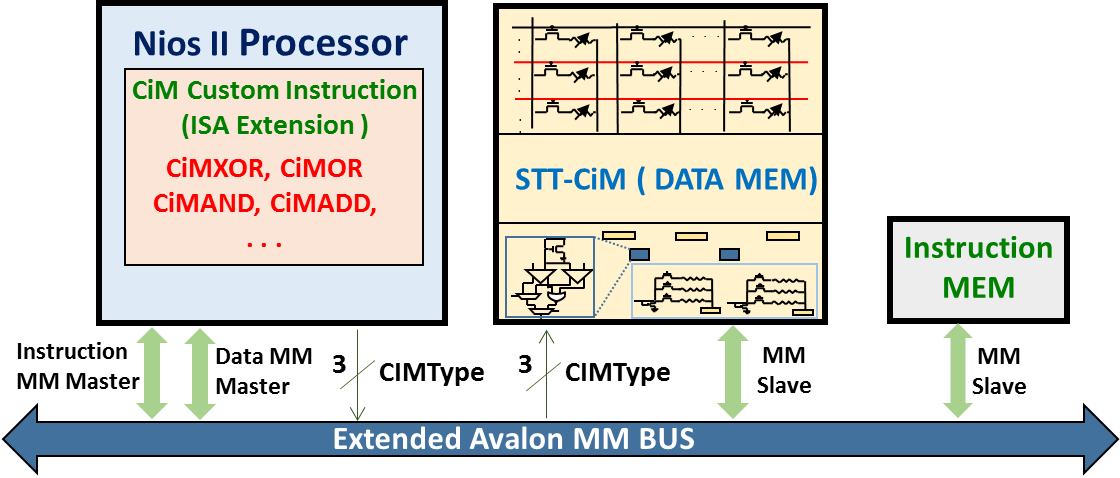}
  \vspace*{-8pt}
  \caption{ System level integration of STT-CiM }
  \label{fig:cim_system}
  \vspace*{-6pt}
\end{figure}

{\bf \noindent Device/Circuit modeling.} We first characterize the bit-cells using SPICE-compatible MTJ models that are based on self-consistent solution of Landau-Lifshitz-Gilbert (LLG) magnetization dynamics and Non-Equilibrium-Green's Function (NEGF) electron transport~\cite{mtj_nanohub}. Table~\ref{tab:deviceparam} shows the MTJ device parameters~\cite{ikeda2010} used in our experiments. Using 45nm bulk CMOS technology and the MTJ models, the memory array along with the associated peripherals and extracted parasitics was simulated in SPICE for read, write and CiM operations to obtain array-level timing and energy characteristics. The obtained characteristics were then used as technology parameters in a modified version of CACTI~\cite{cacti6} that is capable of estimating system-level properties for a spin-based memory. The variation analysis to compute failure rates was performed considering variations in MTJ oxide thickness ($\sigma/\mu$ = 2 \%), transistor $V_{T}$ ($\sigma/\mu$ = 5\%), and MTJ cross sectional area ($\sigma/\mu$ = 5\%).

{\bf \noindent System level simulation.} We evaluated STT-CiM as a 1MB scratchpad for an Intel Nios II processor~\cite{nios2}. Figure~\ref{fig:cim_system} shows the integration of STT-CiM in the memory hierarchy of the programmable processor. In order to expose the STT-CiM operations to software, we extended the Nios II processor's instruction set with custom instructions. The Avalon on-chip bus was also extended to support CiM operations. Cycle-accurate RTL simulation was used to obtain the execution time and the memory access traces for various benchmarks. These traces along with the energy results obtained through the modified CACTI tool were used to estimate the total memory energy. 

{\bf \noindent Benchmark applications.} We evaluate STT-CiM on a suite of twelve algorithms drawn
from various applications (Table~\ref{tab:cim_benchmark}).

\vspace*{-6pt}
\newcolumntype{M}[1]{>{\centering\arraybackslash}m{#1}}
\renewcommand{\arraystretch}{1.2}
\renewcommand{\tabcolsep}{12pt}
\taburulecolor{black}
\begin{table}[hbtp]
  \vspace*{-0pt}
  \centering
  \caption{Device parameters }
  \vspace*{-0pt}
  \begin{tabular}{|c|c|}
	\hline
Material System    &Ta/CoFeB/MgO    \\ \hline 
MTJ Type &PMA \\ \hline 
Saturation Magnetization($M_{S}$)     &1.58T    \\  \hline
Damping Factor, ($\alpha$)   &0.028     \\ \hline
Polarization  &0.62           \\ \hline
Interface Anisotropy  &1.3mJ/$m^{2}$ \\ \hline
MTJ Dimension  &40nm x 40nm x 1.32nm \\ \hline
Oxide Thickness ($t_{ox}$) &1.1nm \\ \hline
Energy Barrier  &65KT \\ \hline
T   &300K \\ \hline
RA Product   &18ohm-$\mu$$m^{2}$ \\ \hline
TMR  &124\% \\ \hline
CMOS Technology &45nm Bulk CMOS \\ \hline
Assumed Variation ($\sigma/\mu$) & $t_{ox}$ = 2\%, MTJ Area = 5\% \\
 &transistor $V_{T}$=5\% \\ \hline
    \end{tabular}    
  \vspace*{-0pt} 
  \label{tab:deviceparam}
\end{table}
\vspace*{-0pt}

\begin{table}[htb]
  \centering 
  \vspace*{-0pt}
  \caption{ Benchmark applications }
  \includegraphics[width=\columnwidth]{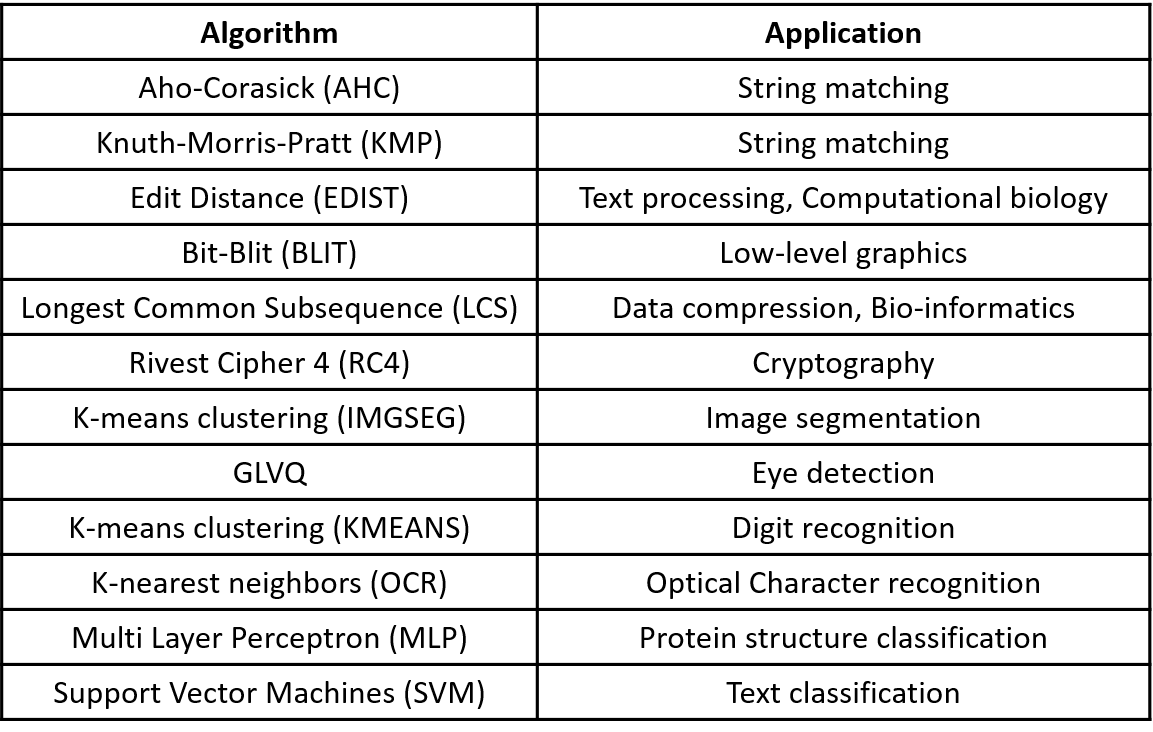}
  \vspace*{-6pt}
  \label{tab:cim_benchmark}
  \vspace*{-6pt}
\end{table}

\vspace*{-0pt}
\section{Results}
\label{sec:results}

\noindent In this section, we first present an array-level analysis of STT-CiM and then quantify its benefits through system-level energy and performance evaluation.

\begin{figure*}[htb]
  \vspace*{-0pt}
  \includegraphics[width=\textwidth]{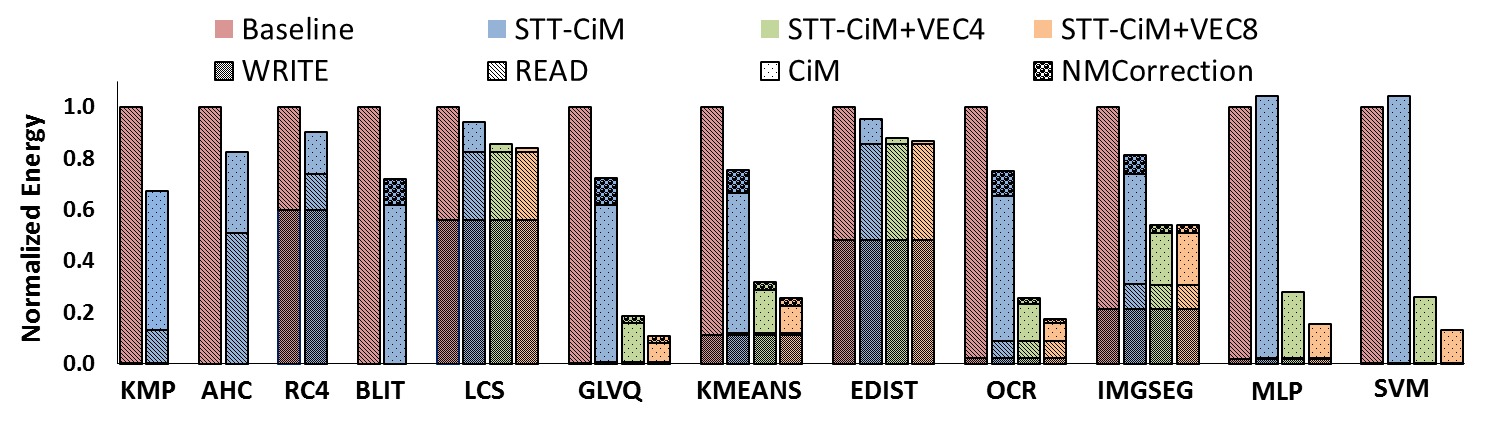}
  \vspace*{-20pt}
  \caption{ Application-level memory energy}
  \label{fig:cim_energyapp_result}
  \vspace*{-6pt}
\end{figure*}

\subsection{Array-level analysis}
\label{subsec:array_level_result}

\begin{figure}[htb]
  \vspace*{-0pt}
  \includegraphics[width=1.0\columnwidth]{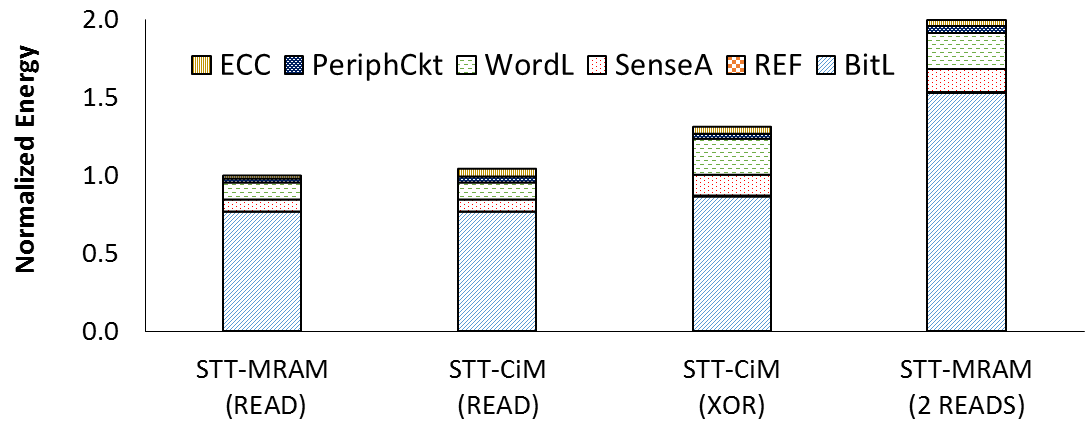}
  \vspace*{-12pt}
  \caption{ Array-level energy evaluation of STT-CiM}
  \label{fig:cimarray_energy_result}
  \vspace*{-4pt}  
\end{figure}

{\bf\noindent Energy.} The second and third bars in Figure~\ref{fig:cimarray_energy_result} show the energy consumed by a standard read operation and a representative CiM operation (CiMXOR) in a 1MB STT-CiM array. Each bar shows the energy breakdown into the major components, \emph{i.e.}, peripheral circuitry (PeriphCkt), wordline (WordL), bitline (BitL), reference generation circuitry (REF), sense amplifier (SenseA), and error correction circuitry (ECC). For comparison, we provide the read energy for an STT-MRAM array of the same capacity (first bar in Figure~\ref{fig:cimarray_energy_result}) and all energy numbers are normalized to this value. A normal read operation in STT-CiM incurs an energy overhead of about 4.4$\%$, which arises primarily due to the extra peripheral circuits (PeriphCkt) and stronger ECC. STT-CiM uses a 3EC4ED ECC scheme (as compared to SECDED in the baseline STT-MRAM), which accounts for about 3$\%$ of the 4.4$\%$ energy overhead. The CiMXOR operation consumes higher energy than a standard read operation mainly due to the charging of multiple wordlines and a slightly higher source line current. However, since a CiM operation replaces {\em two} normal read operations, we also present the energy required for two reads in a standard STT-MRAM (last bar in Figure~\ref{fig:cimarray_energy_result}).
Note that an array-level comparison greatly understates the benefits of STT-CiM, since it does not consider the system-level impact of reduced data transfers between the processor and memory (system-level evaluation is presented in the next subsection). Nevertheless, it is worth noting that even at the array level, STT-CiM consumes 34.2\% less energy than STT-MRAM. The benefits mainly arise from a lower bitline dynamic energy (BitL), since only a single access to the memory array is required for STT-CiM.

\begin{figure}[htb]
  \vspace*{-6pt}
  \includegraphics[width=1.0\columnwidth]{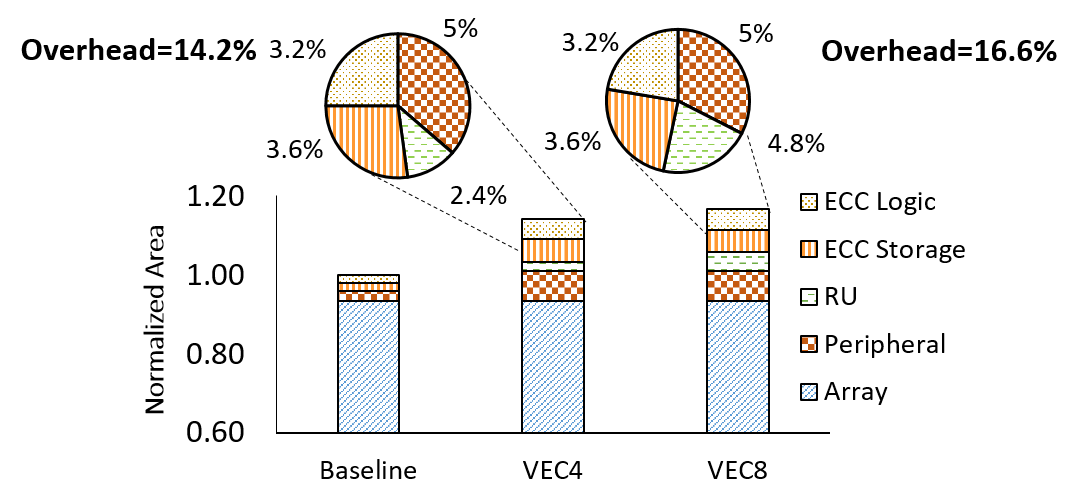}
  \vspace*{-10pt}
  \caption{ Array-level area evaluation of STT-CiM}
  \label{fig:cimarray_area}
  \vspace*{-8pt}  
\end{figure}

{\bf\noindent Area and access time.} Figure~\ref{fig:cimarray_area} shows the area breakdown for two STT-CiM designs that support vector operations of length 4 (VEC4) and 8 (VEC8). As compared to the STT-MRAM baseline, the area overheads for VEC4 and VEC8 are 14.2\% and 16.6\%, respectively. As shown in Figure~\ref{fig:cimarray_area}, Peripheral circuits, ECC storage and ECC Logic are the causes of area overheads (5\%, 3.6\% and 3.2\% respectively). Peripheral circuits include the enhanced address decoder (1.8\%), sense amplifier (0.9\%), and reduce unit (2.3\%). Note that the total area is still dominated by the core array, which remains unchanged. Finally, the access time overhead for STT-CiM was found to be only $\sim$0.8\%, because the wordline and bitline delays dominate the total memory access latency.

\subsection{Application-level memory energy}
\label{subsec:app_energy}

We next present the system level memory energy benefits of using STT-CiM in the programmable processor based system described in Figure~\ref{fig:cim_system}. We evaluated the total memory energy consumed by STT-CiM across the application benchmarks, and compared it with a baseline design that uses standard STT-MRAM. Figure~\ref{fig:cim_energyapp_result} shows the breakdown of different energy components, \emph{viz.} Read, Write and CiM, that contribute to the overall memory energy in both the STT-MRAM and proposed STT-CiM designs. In addition, it also shows the energy overheads due to near memory corrections (NMCorrections) on failing CiM operations. The total memory energy for an application is normalized to the memory energy consumed by the baseline design. For the proposed STT-CiM design, we evaluated a version without vector operations (STT-CiM), and two versions with vector lengths of 4 and 8 (STT-CiM+VEC4 and STT-CiM+VEC8, respectively). Across all benchmarks, 
we observe 1.26x, 2.77x and 3.83x average improvement in energy for STT-CiM, STT-CiM+VEC4 and STT-CiM+VEC8, respectively. 

\begin{figure*}[htb]
  \vspace*{-5pt}
  \centering
  \includegraphics[width=1\textwidth]{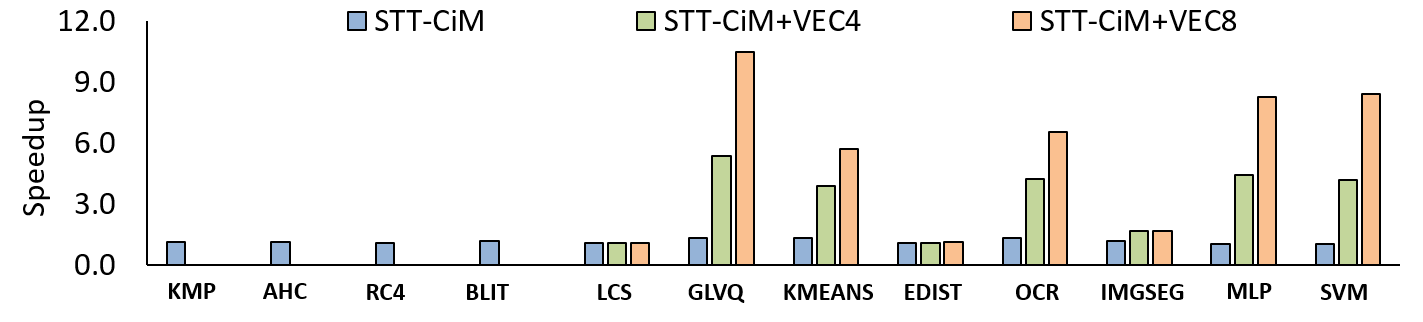}
  \vspace*{-15pt}
  \caption{ Application-level system performance}
  \label{fig:cim_perfapp_result}
  \vspace*{-10pt}
\end{figure*}

\begin{figure}[htb]
  \vspace*{-5pt}
  \includegraphics[width=1.0\columnwidth]{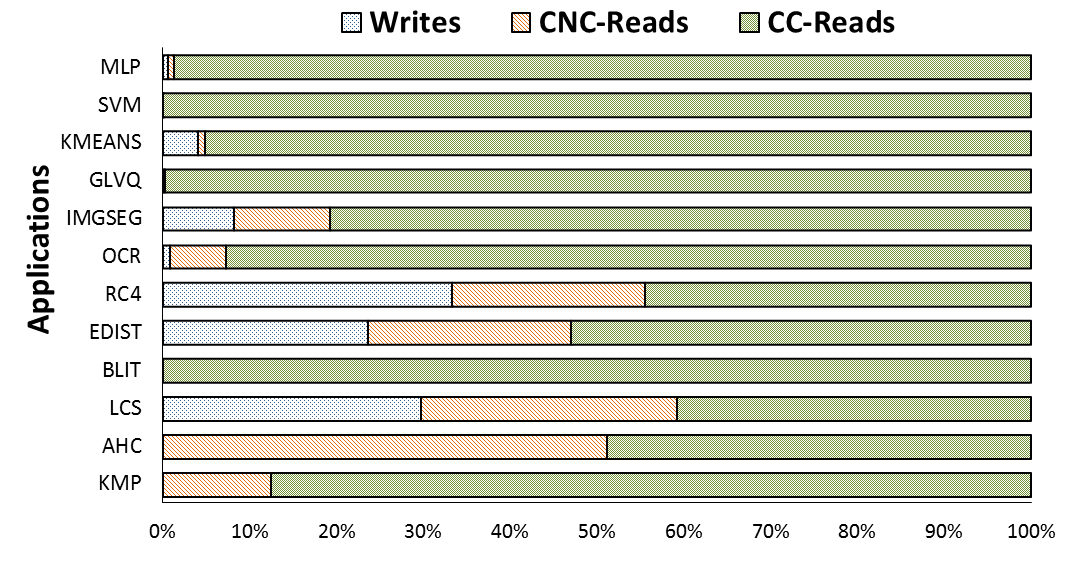}
  \vspace*{-12pt}
  \caption{ Memory access breakdown}
  \label{fig:cim_memory_access}
  \vspace*{-8pt}
\end{figure}

To provide further insights into the energy benefits, Figure~\ref{fig:cim_memory_access} presents a breakdown for memory accesses made by each application into 3 categories --- writes, reads that cannot be converted into CiM operations (CiM non-convertible reads or CNC-Reads), and CiM convertible reads (CC-Reads). We see that applications where CC-Reads dominate the total memory accesses (KMP, BLIT, GLVQ, KMEANS, OCR, IMGSEG, MLP, SVM in Figure~\ref{fig:cim_memory_access}) experience higher energy benefits from STT-CiM (Figure~\ref{fig:cim_energyapp_result}). Among these applications, those that benefit from vectorization achieve the highest savings (GLVQ, KMEANS, OCR, MLP, SVM). Applications with relatively fewer CC-Reads or more frequent writes (AHC, LCS, RC4, EDIST) exhibit relatively lower energy savings. CNC-Reads and writes are not benefited by STT-CiM, and writes in particular consume 
significantly ($\sim$3x) higher energy than reads. The energy overheads due to additional writes incurred for data alignment in Type II and Type III compute patterns was observed to be 0.8\% and 0.3\%, respectively.

\subsection{System-level performance}
\label{subsec:app_performance}

Figure~\ref{fig:cim_perfapp_result} shows the speedup for the Nios II processor system integrated with STT-CiM across various applications. The speedup shown in the figure is with respect to the baseline design, \emph{i.e.}, the processor system integrated with a standard STT-MRAM based memory. As discussed in Section~\ref{subsec:archi_isa}, CiM lowers the total number of memory accesses as well as the number of instructions executed, which leads to performance benefits at the system level. Overall, for STT-CiM without vector operations, we observe performance benefits ranging from 1.07X to 1.36X.
With vector operations, the average speedup increased to 3.25x and 3.93x for vector lengths of 4 and 8, respectively. 
Comparing Figures~\ref{fig:cim_memory_access} and~\ref{fig:cim_perfapp_result}, we see that the factors that indicate higher energy savings for an application (large fraction of memory accesses are CC-Reads, opportunities for vectorization exist) are also predictive of higher performance improvements.

\begin{figure}[htb]
  \vspace*{-6pt}
  \includegraphics[width=1\columnwidth]{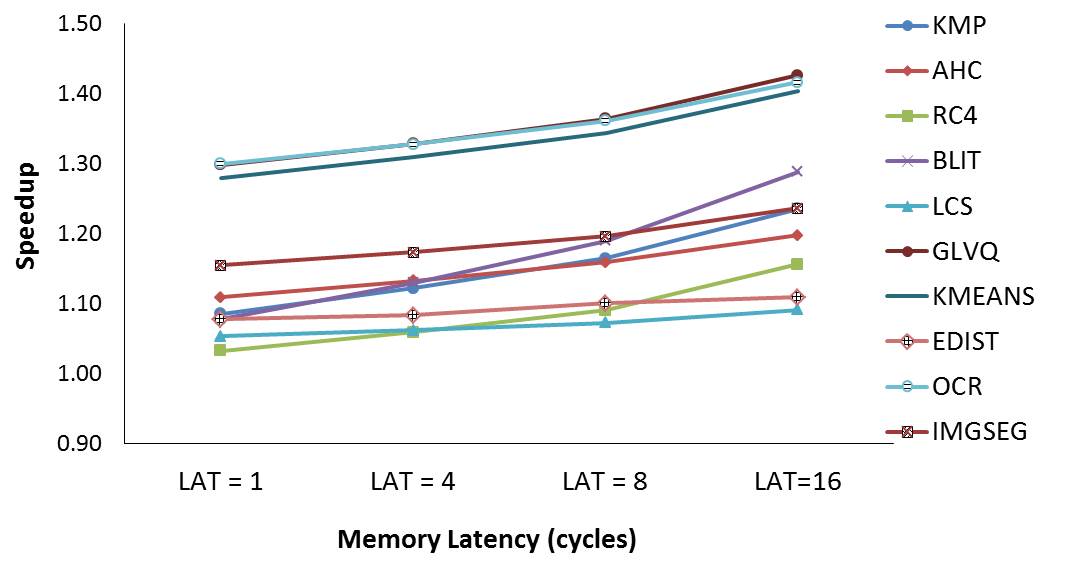}
  \vspace*{-12pt}
  \caption{ Performance sensitivity to memory latency }
  \label{fig:cim_lat_result}
  \vspace*{-6pt}
\end{figure}

In order to demonstrate the performance sensitivity to memory latency, we vary the memory latency and evaluate the execution time for each application. Figure~\ref{fig:cim_lat_result} shows the results of this sensitivity analysis. On the Y-axis, we have the speedup of STT-CiM over STT-MRAM, and on the X-axis the memory latency. We observe that STT-CiM yields higher performance benefits at higher memory latency. This is attributed to the fact that the reduced number of memory accesses for STT-CiM has a larger impact on system performance. On an average, we achieve 1.13x speedup for a memory latency of 1 cycle, and 1.26x speedup for a memory latency of 16 cycles, thereby illustrating the effectiveness of the proposed approach.

\vspace*{-0pt}
\section{Conclusion}
\label{sec:conclusion}
{\bf \noindent} STT-MRAM is a promising candidate for future on-chip memories. In this work, we proposed STT-CiM, an enhanced STT-MRAM that can perform a range of arithmetic, logic and vector compute-in-memory operations. We addressed a key challenge associated with these in-memory operations, \emph{i.e.} reliable computation under process variations. We utilized the proposed design (STT-CiM) as a scratchpad in the memory hierarchy of a programmable processor, and introduced ISA extensions and on-chip bus enhancements to support in-memory computations. We proposed architectural optimizations and data mapping techniques to enhance the efficiency of STT-CiM. A device-to-architecture simulation framework was used to evaluate the benefits of STT-CiM. Our experiments indicate that STT-CiM achieves substantial improvements in energy and performance, and shows considerable promise in alleviating the processor-memory gap.

\vspace*{-0pt}
\scriptsize
\bibliographystyle{unsrt}
\bibliography{references}
\vspace*{-30pt}
\begin{IEEEbiography}
[{\includegraphics[width=1.1in,height=1.25in,clip,keepaspectratio]{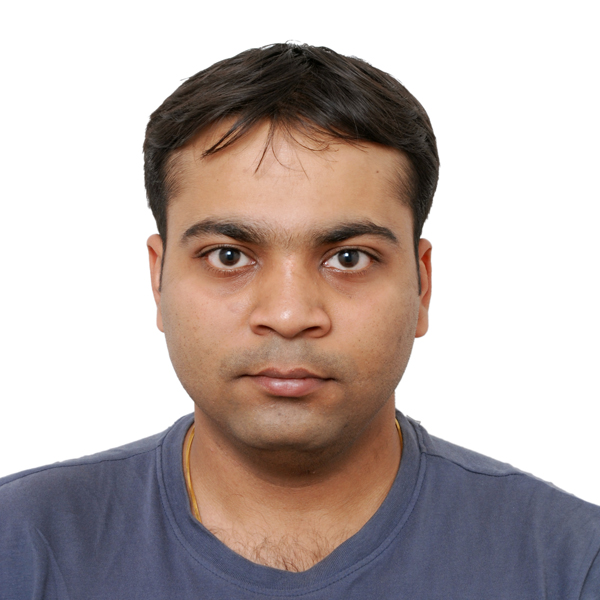}}]
{Shubham Jain} is currently a PhD student in the School of Electrical and Computer Engineering, Purdue University. His research interests include exploring circuit and architectural techniques for emerging post-CMOS devices and computing paradigms such as spintronics, approximate computing and neuromorphic computing. He has a B.Tech(Hons.) degree in Electronics and Electrical Communication Engineering from the Indian Institute of Technology, Kharagpur, India, in 2012. After graduation, he worked for two years in Qualcomm, Bangalore, India. He is a recipient of the Andrews Fellowship from Purdue University, in 2014. 
\end{IEEEbiography}
\vspace*{-30pt} 
\begin{IEEEbiography}
[{\includegraphics[width=1.1in,height=1.25in,clip,keepaspectratio]{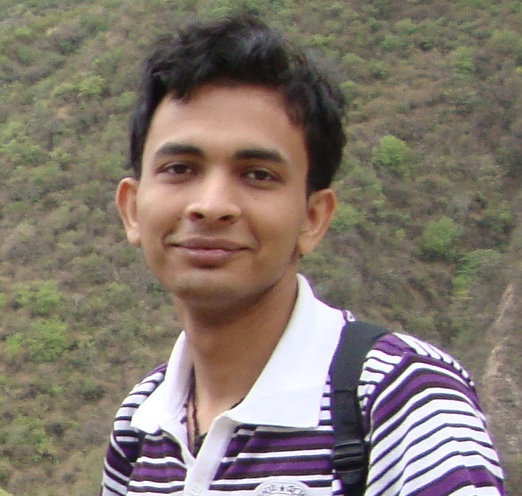}}]
{Ashish Ranjan} received the BTech degree in electronics engineering from the Indian Institute of Technology (BHU), Varanasi, India, in 2009. He is currently working towards a PhD degree in the School of Electrical and Computer Engineering, Purdue University, West Lafayette, IN. His industry experience includes three years as a senior member technical staff in the Design Creation Division, Mentor Graphics Corporation, Noida, India. His primary research interests include circuit-architecture codesign for emerging technologies and approximate computing. He was awarded the University Gold Medal for his academic performance by IIT (BHU), Varanasi in 2009. He also received the Andrews Fellowship from Purdue University in 2012. 
\end{IEEEbiography}
\vspace*{-20pt}
\begin{IEEEbiography}
[{\includegraphics[width=1.1in,height=1.25in,clip,keepaspectratio]{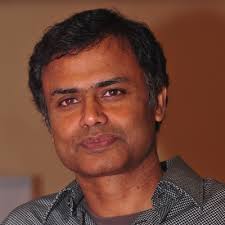}}]
{Kaushik Roy} received the BTech degree in electronics
and electrical communications engineering
from the Indian Institute of Technology, Kharagpur,
India, and the PhD degree from the Department
of Electrical and Computer Engineering,
University of Illinois at Urbana-Champaign in
1990. He was with the Semiconductor Process
and Design Center of Texas Instruments, Dallas,
where he worked on FPGA architecture development
and low-power circuit design. He joined the
electrical and computer engineering faculty at Purdue
University, West Lafayette, IN, in 1993, where he is currently Edward
G. Tiedemann Jr. Distinguished Professor. His research interests include
spintronics, device-circuit co-design for nano-scale Silicon and non-Silicon
technologies, low-power electronics for portable computing and wireless
communications, and new computing models enabled by emerging
technologies. He has published more than 600 papers in refereed journals
and conferences, holds 15 patents, graduated 60 PhD students, and
is coauthor of two books on Low Power CMOS VLSI Design (Wiley \&
McGraw Hill). He received the US National Science Foundation Career
Development Award in 1995, IBM faculty partnership award, ATT/Lucent
Foundation award, 2005 SRC Technical Excellence Award, SRC Inventors
Award, Purdue College of Engineering Research Excellence Award,
Humboldt Research Award in 2010, 2010 IEEE Circuits and Systems
Society Technical Achievement Award, Distinguished Alumnus Award
from Indian Institute of Technology, Kharagpur, Fulbright-Nehru Distinguished
Chair, and Best Paper Awards at 1997 International Test Conference,
IEEE 2000 International Symposium on Quality of IC Design, 2003
IEEE Latin American Test Workshop, 2003 IEEE Nano, 2004 IEEE International
Conference on Computer Design, 2006 IEEE/ACM International
Symposium on Low Power Electronics \& Design, and 2005 IEEE Circuits
and System Society Outstanding Young Author Award (Chris Kim), 2006
IEEE Transactions on VLSI Systems Best Paper Award, 2012 ACM/IEEE
International Symposium on Low Power Electronics and Design Best
Paper Award, 2013 IEEE Transactions on VLSI Best Paper Award. He
was a Purdue University Faculty scholar (1998-2003). He was a
Research Visionary board member of Motorola Labs (2002) and held the
M.K. Gandhi Distinguished Visiting faculty at Indian Institute of Technology
(Bombay). He has been in the editorial board of IEEE Design and
Test, IEEE Transactions on Circuits and Systems, IEEE Transactions on
VLSI Systems, and IEEE Transactions on Electron Devices. He was the
guest editor for Special Issue on Low-Power VLSI in the IEEE Design and
Test (1994) and IEEE Transactions on VLSI Systems (June 2000), IEE
Proceedings—Computers and Digital Techniques (July 2002), and IEEE
Journal on Emerging and Selected Topics in Circuits and Systems
(2011). He is a fellow of the IEEE
\end{IEEEbiography}

\vspace*{-20pt}
\begin{IEEEbiography}
[{\includegraphics[width=1.1in,height=1.25in,clip,keepaspectratio]{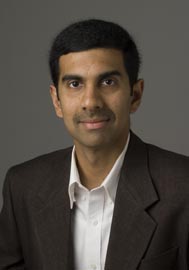}}]
{Anand Ragunathan}
is a Professor of Electrical and Computer Engineering and Chair of the VLSI area at Purdue University, where he directs research in the Integrated Systems Laboratory. His current areas of research include domain-specific architecture, system-on-chip design, computing with post-CMOS devices, and heterogeneous parallel computing. Previously, he was a Senior Research Staff Member at NEC Laboratories America, where he led projects on system-on-chip architecture and design methodology. He has also held the Gopalakrishnan Visiting Chair in the Department of Computer Science and Engineering at the Indian Institute of Technology, Madras. 

Prof. Raghunathan has co-authored a book, eight book chapters, and over 200 refereed journal and conference papers, and holds 21 U.S patents. His publications received eight best paper awards and five best paper nominations. He received a Patent of the Year Award and two Technology Commercialization Awards from NEC, and was chosen among the MIT TR35 (top 35 innovators under 35 years across various disciplines of science and technology) in 2006.

Prof. Raghunathan has been a member of the technical program and organizing committees of several leading conferences and workshops, chaired premier IEEE/ACM conferences (CASES, ISLPED, VTS, and VLSI Design), and served on the editorial boards of various IEEE and ACM journals in his areas of interest. He received the IEEE Meritorious Service Award and Outstanding Service Award. He is a Fellow of the IEEE and Golden Core Member of the IEEE Computer Society. Prof. Raghunathan received the B. Tech. degree from the Indian Institute of Technology, Madras, and the M.A. and Ph.D. degrees from Princeton University.
\end{IEEEbiography}

\end{document}